%% file: ms.tex
\documentclass[sigconf,usenames,dvipsnames,table]{acmart}
\usepackage{filecontents}
\include{macro}
\includecomment{submission}
\excludecomment{cameraready}

\begin{document}
%-------------------------------------------------------------------------------

%don't want date printed
\date{}

%\title{\Large \bf Enabling Data-Driven Security Analysis of Home
%Automation} \title{Enabling a Natural Perspective for Home Automation
%Security Analysis} \title{\tool: Enabling Using Natural Home Automation
%Analysis}
\title{\tool: Enabling a Natural Perspective of Home Automation}
\settopmatter{printfolios=true,printccs=false}

%\title{A Study of Data Store-based Home Automation}
\author{{Sunil Manandhar, Kevin Moran, Kaushal Kafle, Ruhao Tang, Denys Poshyvanyk, Adwait Nadkarni}}
\affiliation{William \& Mary, Williamsburg, VA, USA}
\email{{smanandhar, kpmoran, kkafle, rtang, denys, nadkarni}@cs.wm.edu}
\date{}

%\author{{\rm Sunil Manandhar, Kevin Moran, Kaushal Kafle, Ruhao Tang,
%Denys Poshyvanyk, Adwait Nadkarni}}
%\affiliation{William \& Mary, Williamsburg, VA, USA}
%\email{{sunil, kpmoran, kkafle, rtang, denys, nadkarni}@cs.wm.edu}
%for single author (just remove % characters)
%\author{
%{\rm Anonymous Authors}\\
% copy the following lines to add more authors
% \and
% {\rm Name}\\
%Name Institution
%} % end author

\input{abstract}

\maketitle

% Paper Body
%-------------------------------------------------------------------------------
\input{intro}
\input{motivation}

\input{background}

\input{design}

\input{data_collection_insights}

\input{naturalness}
\input{validity}

\input{case-studies}
\input{related-work}
\input{discussion}

%\input{conclusion}
%-------------------------------------------------------------------------------

\begin{cameraready}
%-------------------------------------------------------------------------------
\section*{Acknowledgments}
%-------------------------------------------------------------------------------

The USENIX latex style is old and very tired, which is why
there's no \textbackslash{}acks command for you to use when
acknowledging. Sorry.

%-------------------------------------------------------------------------------
\section*{Availability}
%-------------------------------------------------------------------------------

On publication, we plan to release the anonymized data used for training \tool as well as the code.
\end{cameraready}

%-------------------------------------------------------------------------------
\bibliographystyle{ACM-Reference-Format}
%\bibliography{bibliographies/os,bibliographies/semeru,bibliographies/iot,bibliographies/misc,bibliographies/phone,bibliographies/monkeylab}
\bibliography{ms}
%\newpage
\appendix
\input{survey_methodology}

\input{policies}

%%%%%%%%%%%%%%%%%%%%%%%%%%%%%%%%%%%%%%%%%%%%%%%%%%%%%%%%%%%%%%%%%%%%%%%%%%%%%%%%
\end{document}

%% file: macro.tex
\usepackage{epsf}
\usepackage{epsfig}
\usepackage{graphicx}
\usepackage{times}
\usepackage{ifthen}
\usepackage{array}
\usepackage{comment}
\usepackage{url}
\usepackage{cellspace}
\usepackage{latexsym}
\usepackage{amsfonts}
\usepackage{subfigure}
\usepackage{multirow}
\usepackage{multicol}
\usepackage{tabularx}
\usepackage{listings}
\usepackage{amssymb}
\usepackage{mathtools}
\usepackage{wrapfig}
\usepackage{xspace}
\usepackage{enumitem}
\usepackage[compact]{titlesec}
\usepackage{footnote}
\usepackage{gensymb}
\usepackage{todonotes}
\usepackage{tabularx}
\usepackage{makecell}
\newboolean{showcomments}

\usepackage[utf8]{inputenc}
\usepackage{tikz}
\usepackage{pgfplots,pgfplotstable}\pgfplotsset{compat=1.9}
\usetikzlibrary{external}
\setboolean{showcomments}{true}

\ifthenelse{\boolean{showcomments}}
  {\newcommand{\nb}[2]{
    \fbox{\bfseries\sffamily\scriptsize#1}
    {\sf\small$\blacktriangleright$\textit{#2}$\blacktriangleleft$}
   }
   
  }
  {\newcommand{\nb}[2]{}
   
  }

\newcommand\ADWAIT[1]{{\color{red} \nb{ADWAIT}{#1}}}

\newcommand\TODO[1]{{\color{red} \nb{TODO}{#1}}}

\newcommand{\tool}{{H$\epsilon$lion}\xspace}
\newcommand{\tools}{{H$\epsilon$lion}'s\xspace}

\newcommand{\xy}{{\color{red} \textbf{XY}}\xspace}

\newcommand{\ie}{\textit{i.e.},\xspace}
\newcommand{\eg}{\textit{e.g.},\xspace}

\newcommand{\fnumber}[1]{{$\mathcal{F}_{#1}$}}
\newcommand{\resq}[1]{\textbf{$RQ_{#1}$}}
\newcommand{\finding}[2]{\vspace{0.25em}\noindent{\bf {\em Finding\ {#1}} (\fnumber{{#1}}): {#2}.}}

\newcommand{\nest}{{NEST}\xspace}

\newcommand{\smartthings}{{SmartThings}\xspace}
\newcommand{\ngram}{{n-gram}\xspace}
\newcommand{\ngrams}{{n-grams}\xspace}

\newcommand{\sequence}[1]{{\sf \small #1}}
\newcommand{\routine}[2]{{\bf IF} {\sf \small #1} {\bf THEN} {\sf \small #2}}

\titlespacing{\section}{0pt}{2ex}{1ex}
\titlespacing{\subsection}{0pt}{1ex}{0ex}
\titlespacing{\subsubsection}{0pt}{0.5ex}{0ex}

\newcounter{taskctr}

\newcommand{\myparagraph}[1]{\vspace{0.5em}\noindent{\bf #1}:}

\definecolor{dkgreen}{rgb}{0,0.6,0}
\definecolor{gray}{rgb}{0.5,0.5,0.5}
\definecolor{mauve}{rgb}{0.58,0,0.82}
\definecolor{shadecolor}{rgb}{0.95,0.95,0.95}
\setlist[itemize]{leftmargin=*, noitemsep, topsep=1pt}
\setlist[enumerate]{leftmargin=*, noitemsep, topsep=1pt}

\makeatletter
\renewcommand{\paragraph}{%
	\@startsection{paragraph}{4}%
	{\z@}{0.5ex \@plus 0ex \@minus .2ex}{-1em}%
	{\normalfont\normalsize\bfseries}%
}
\makeatother

\makeatletter
\g@addto@macro\normalsize{%
	\setlength\abovedisplayskip{0pt}
	\setlength\belowdisplayskip{0pt}
	\setlength\abovedisplayshortskip{-10pt}
	\setlength\belowdisplayshortskip{0pt}
}
\makeatother

\DeclareGraphicsExtensions{.pdf,.png}
\graphicspath{{img/}}

\renewcommand{\refname}{{\hskip 0pt plus 0.1fil minus 0pt} References Cited\hfil}                   %%

%% file: abstract.tex
%-------------------------------------------------------------------------------
\begin{abstract}
%-------------------------------------------------------------------------------
Security researchers have recently discovered significant security and safety issues related to home automation, and developed approaches to address them.
However, such approaches often face design and evaluation challenges which arise from their restricted perspective of home automation that is bounded by the IoT apps they analyze.
The challenges of past work can be overcome by relying on a deeper understanding of {\em realistic} home automation usage.
More specifically, the availability of {\em natural home automation scenarios}, \ie sequences of home automation events that may realistically occur in an end-user's home, could help security researchers design better security/safety systems.
%, and more specifically, a perspective of home automation derived from {\em user-driven} routines, \ie routines that are increasingly being created by users through intuitive UIs provided by most major platforms.

%However, t
%Security and safety research for protecting smart homes could be significantly enhanced with representative home automation usage, \ie of {\em natural}, home automation {\em scenarios} that may realistically occur in an end-user's home.
%For example, security researchers often have to imagine use/misuse cases when specifying security and safety policies; automatically-generated scenarios could replace the use-cases, significantly reducing the burden on the researchers. 
%which is challenging without any intuition of real use cases from end-user homes.

This paper presents \tool, a framework for building a natural perspective of home automation.
\tool identifies the regularities in user-driven home automation, \ie automation that is built from {\em user-driven} routines that are increasingly being created by users through intuitive platform UIs. 
%To address this need for user-driven data in smart home security tools, this paper presents \tool, a framework for identifying and modeling the regularities in user-driven home automation, and predicting {\em natural} home automation {\em scenarios}.
%, \ie sequences of smart home events that may realistically occur in an end-user's home.
%To initialize \tool, we consider {\em user-driven} routines, \ie trigger-action programs created by end-users using the user-interfaces provided by 
%Many modern smart home platforms increasingly enable end-users to easily create trigger-action routines themselves, \ie such {\em user-driven} routines are directly representative of end-user requirements.
%\tool leverages user-driven routines, along with the following ``naturalness hypothesis'', to model the regularities in home automation using existing statistical modeling techniques: while the theoretical scope of home automation event sequences caused by routines may seem exponentially large and variable, {\em in practice}, routines are created by humans with specific intentions, and hence, exhibit {\em patterns} that make them {\em predictable}.
%\tool consists of an end-to-end framework for collecting, representing, modeling, and generating realistic smart home events through a novel adaption of statistical n-gram language modeling. 
Our intuition for designing \tool is that smart home event sequences created by users exhibit an inherent set of semantic patterns, or ``\textit{naturalness}'' that can be modeled and used to generate valid and useful scenarios. 
To evaluate our approach, we first empirically demonstrate that this ``naturalness'' hypothesis holds, with a corpus of 30,518 home automation events, constructed from 273 routines collected from 40 users.
We then demonstrate that the scenarios generated by \tool are reasonable and valid from the end-user perspective, through an evaluation with 16 external evaluators.
We further demonstrate the usefulness of \tools scenarios by addressing the challenge of generating security/safety policies for home automation, and specifying 17 security policies with significantly less effort than existing approaches. 
%We then apply \tool to the task of generating and specifying security and safety policies for smart homes, and illustrate its effectiveness and potential to augment existing policy enforcement techniques. 
We conclude by discussing key takeaways and future research challenges enabled by \tools natural perspective of home automation.
%Finally, we evaluate the validity of the scenarios generated by \tool with \xy external evaluators, demonstrating significant perceived naturalness, while also uncovering critical insights and design challenges in this exciting new research direction.
\end{abstract}

%% file: intro.tex
\section{Introduction}
\label{sec:intro}

Smart home technology has reached a critical mass, with the global smart home market projected to grow to over \$53 billion by the year 2022~\cite{home-adoption}.
%it's popularity directly driven by the consumer demand for home automation.
%Smart home products have become extremely popular with consumers, and over 20 billion of these devices are projected to be in active use by 2020~\cite{iot-adoption}.
This market is primarily driven by the consumer demand for seamless home automation, which not only includes the ability to control smart home devices remotely, but also the flexibility of having devices react to changes in the state of the user or the home.
Popular smart home platforms such as Samsung \smartthings~\cite{smartthingsroutines} and Google~\nest~\cite{nestdoc} enable automation through trigger-action programs known as {\em routines}, which help users induce {\em action} events in their smart home when a specific {\em trigger} condition is satisfied; \eg when the user gets home (\ie trigger) turn the security camera OFF (\ie action).
Routines are the building blocks of home automation.

Prior research has often studied routines, and more specifically, {\em IoT apps} published in marketplaces such as the Samsung SmartThings Official Repository~\cite{smartapp-repo}, to explore the security, safety, and privacy properties of home automation ~\cite{tzl+17,jcw+17,cmt18,whbg18,nsq+18,nsq+18,dh18,cbs+18,ctm19,kmm+19}.
For instance, researchers have analyzed routines to detect the security and safety consequences of combinations of routines (\eg~Soteria~\cite{cmt18}, IoTGuard~\cite{ctm19}, and IoTMon~\cite{dh18}), enable contextual integrity~(\eg ContexIoT~\cite{jcw+17}), ensure informed user-consent (\eg SmartAuth~\cite{tzl+17}), provide general provenance information (\eg~ProvThings~\cite{whbg18}), and track privacy leaks (\eg~Saint~\cite{cbs+18}).

While prior research gives us a useful estimate of problems that could occur in smart homes, %its heavy focus on IoT apps means that 
 the findings would only be actionable {\em iff} users utilize a specific combination of these apps in a specific manner (\eg triggered in a particular order).
However, without any complementary insight into what routines people \textit{actually} deploy, or how they execute them, it is difficult to put the findings of prior work into perspective, or to prioritize them. 
Generally speaking, without observing realistic home automation usage, it is difficult to practically instantiate and evaluate research in this domain. 
For instance, a researcher creating policies for Soteria~\cite{cmt18} or IoTGuard~\cite{ctm19} needs to come up with use/misuse scenarios for smart home devices, which requires significant manual effort, is limited by the perspective of the researcher, and may not reflect the use/misuse scenarios that occur in user homes.
Similarly, without access to actual events that occur in user homes, systems like ContextIoT~\cite{jcw+17} and IoTSAN~\cite{nsq+18} would continue to be evaluated with random events as input, which may not reflect the practical performance of the system.

From these instances, we observe a common gap that limits the practicality of current research: the lack of {\em natural home automation scenarios}, \ie sequences of events that would be {\em reasonably likely} to occur in end-user homes.
For example, consider the following sequence of three events:

\begin{center}
{\sf \small temperature drops below 70F $\rightarrow$ electric blanket turns ON \\$\rightarrow$ smoke detector detects smoke and sounds the alarm.}
\end{center}

 This scenario is fairly common, as electric blankets have often led to fires, especially when left ON or unattended for a long time~\cite{blanket-fire-1,blanket-fire-2,blanket-fire-3}. %In fact, increasing automation can only increase the risk of blankets left ON unattended. 
The design and evaluation of security and safety systems for the smart home would stand to significantly benefit if such scenarios were widely available.
This paper is the first step in our vision of {\em developing a natural perspective of home automation} that can be used to predict natural scenarios, to be used for the design and evaluation of security systems.
We explore how this natural perspective can be obtained by learning regularities from user-driven routines in the context of home automation.
%{\sf (1)} leveraging user-driven routines for {\sf (2)} learning the regularities in home automation.
%We now briefly motivate the two core components of our approach for achieving this natural perspective: {\sf (1)} a focus on {\em user-driven routines} and {\sf (2)} learning the regularities in user-driven home automation enabling practical analysis of user-driven routine via the {\em naturalness hypothesis} for home automation.

\textit{User-driven routines} are routines that end-users configure through interactive user interfaces (UIs) provided by platform vendors (\eg~the SmartThings official app~\cite{smartthingsroutines}) and third-party smart home managers (\eg Yeti~\cite{yetiapp} and Yonomi~\cite{yonomiapp}).
Such routines are a realization of the ``end-user programming'' paradigm in smart homes, as users can configure devices and assign triggers and actions without writing a single line of code.
Users are empowered to craft their own routines to fit their workflows, without depending upon developer-defined IoT apps for the desired functionality, \ie user-driven routines {\em directly} reflect the requirements of end-users. 
The following example clearly illustrates our motivation behind focusing on user-driven routines:

\vspace{0.25em}
\centerline{
\routine{the doorbell rings}{turn the security camera ON}
}
\vspace{0.25em}
\noindent The above routine is a straightforward use-case one could imagine with a smart doorbell and a security camera, \ie it ensures that the camera stays OFF for most of the time the user is home (for privacy), but turns ON when significant events happen, such as when someone is at the door.
Indeed, users we collected routines from as a part of our study specified this particular routine (Sec.~\ref{sec:data_collection}). 
Yet, this routine was not found in any of the 187 IoT apps in the public \smartthings repository; in fact, out of the 273 routines (233 unique) created by our 40 users, more than 42.49\% were not represented by any IoT app.
This apparent mismatch between developer-provided functionality and actual user needs is a key motivator behind our focus on user-driven routines for modeling home automation.

%\ADWAIT{Lets save this for the discussion. That is, a nice takeaway from this paper, another finding we can talk about later. That is, this paper motivates this rather open ended research challenge.}
%This phenomenon is in complete contrast to app-based platforms, such as Android, where users are entirely dependent on apps built by third-party developers.
%Such user-driven home automation may introduce a set of largely unexplored security implications that may not be apparent from just analyzing IoT apps. 
%This gap is illustrated by a data collection experiment we describe later, where out of 250 routines created by 37 users, 42.8\% are not represented by any of the IoT apps in the public \smartthings repository.
%This apparent mismatch between developer-provided functionality and user needs illustrates the generally unknown nature of user-driven routines that complicates practical security analyses of home automation, and motivates a \underline{novel research challenge}:%  unique to home automation:
%These arguments motivate a \underline{novel research challenge} that is unique to home automation: 
%\textit{ unlike past research on mobile platforms, characterizing home automation environments solely according to apps published in markets is not sufficient, as users may not necessarily use IoT apps in favor of easily-created user-driven routines}.

We propose a novel approach that uses statistical language modeling~\cite{Manning:1999} to identify the regularities in user-driven home automation, and leverages them to generate scenarios.
Our approach builds upon a key result from the domain of Natural Language Processing (NLP), which states that while a natural language such as English may be extremely expressive in theory, {\em in practice}, the actual range of ``utterances'', \ie the use of the language by people, is ``natural'', \ie generally exhibits certain patterns and is, thus, predictable. 
%That is, while language usage may be highly contextual, it 
This result was successfully leveraged and extended by Hindle {\em et al.}, who demonstrated that source code, just like natural language, is the culmination of human effort, and as a result, contains repetitive patterns making it predictable~\cite{Hindle:ICSE12}. 
We hypothesize that just like natural language and software corpora, home automation usage is also natural, and can be usefully modeled using existing language modeling techniques, to identify regularities that would help us generate reasonably common home automation scenarios.
%Hence, just like natural language corpora, even software can be usefully modeled using existing language modeling techniques.

%We hypothesize that user-driven home automation is natural, just like software or natural language corpora, and can be usefully modeled using existing language modeling techniques, exhibiting regularities that would help us generate reasonably common home automation scenarios.
%We explore the possibility of leveraging this notion of naturalness to effectively evaluate home automation.
Specifically, we define the notion of a {\em home automation sequence}, which is the full, ordered, set of routines that the end-user has scheduled to execute in their smart home (\ie analogous to functions called within a program). 
Using this definition, we frame the following \underline{core} hypothesis to be tested in this paper:
\vspace{1.0em}

 \hspace*{-0.42cm}
\noindent\fbox{%
    \parbox{1.015\linewidth}{%
Home automation sequences created by humans are implicitly \textbf{\em natural}, \ie while they are subject to certain contextual constraints, these sequences exhibit logical patterns, making them predictable. Hence, statistical language modeling can be leveraged to analyze corpora of home automation sequences and predict \textbf{\em useful} scenarios to facilitate the design and evaluation of security systems.
	}
}
\vspace{1.0em}
 
% \ADWAIT{These three sentences have been commented out as they seemed rather unfocused, especially after we have tied the paper around generating scenarios. I feel that they would look really good at the end, in the discussion section.} 
%The practical choice of modeling natural utterances (\ie user-driven routines) rather than all potential theoretical instances of routines could lead to a defining change in our understanding of home automation and its security implications from the user's perspective.
%This paper explores the insights that can be gained by taking a data-driven approach to examining the corpora of user-driven home automation programs. 
%In doing so, we investigate the ability of statistical language models to embody the {\em natural} regularities in such corpora, and further, make {\em useful} predictions about future events in a user's smart home given a history, for a practical security evaluation.
 
 % Helion
We present a framework that enables a natural perspective for Home automation security EvaLuatION (\tool).  
We initialize \tool with routines collected from end-users.
Moreover, we observe that the order in which two routines execute may have different, even contradictory, security implications.
Hence, for a precise characterization, \tool obtains {\em execution indicators}, \ie clues from the end user about the potential execution of individual routines, and performs an {\em informed} scheduling of the routines in the user's home, which results in an ordered {\em home automation sequence} composed of home automation events as they are scheduled to execute in the user's home.
\tool then leverages the \ngram language model to learn patterns from multiple home automation sequences, and predict natural scenarios, \ie sequences of future events that would occur, given the past history of events in a user's home. 
For example, the ``electric blanket'' scenario presented earlier in this section was generated 
%\DENYS{I am still not a big fan of using "predicting". Maybe we should say "generating" since these are generative models?} 
by \tool.
This paper makes the following contributions:
\begin{itemize}
\item {\bf \tool:} We present, \tool, a novel framework designed to model user-driven home automation sequences, using statistical language modeling, to generate natural home automation scenarios.
%We design the \tool framework which constructs home automation sequences from user-driven routines, models them using the \ngram language model, and predicts scenarios, which can then be used to design and evaluate security systems. 
\item {\bf Naturalness of home automation:} We empirically test the naturalness hypothesis over a home automation corpus (called {\sf HOME}) consisting of 30,518 events, split among event sequences from 40 users, created using 273 routines. Our evaluation demonstrates that home automation is indeed {\em natural} and predictable, even more so than natural language and software corpora.
\item {\bf Validity of predicted scenarios:} We study the validity of our scenarios with 16 additional external evaluators. \tools scenarios are generally seen as reasonable/natural by evaluators. 
\item {\bf Usefulness of scenarios:} We demonstrate the usefulness of the scenarios by using \tool to generate security and safety policies for home automation. Our evaluation demonstrates that \tools approach significantly automates the use/misuse case analysis workflow by eliminating the need to imagine scenarios. Our approach enabled us to semi-automatically discover 27 unsafe scenarios in user-driven home automation and specify 17 security policies to address them.
\end{itemize}

Our analysis of the results and feedback from users leads to key findings (\fnumber{1}$\rightarrow$\fnumber{14}) that demonstrate the strengths of \tool and surprising aspects of home automation, as well as additional design challenges and future opportunities in this exciting domain. 

%\myparagraph{Data Availability and IRB Approval} We plan to make anonymized datasets (\ie including the {\sf HOME} corpus) available upon publication. Further, all the user studies and surveys described in this paper were approved by our institutional IRB.
 
%{\color{red} Read until this point}
The rest of the paper is organized as follows: Section~\ref{sec:motivation} motivates the need for natural scenarios, and provides key intuition into language modeling.
Section~\ref{sec:design} describes \tool. 
Section~\ref{sec:data_insights} details our data collection approach and initial insights.
Section~\ref{sec:eval_overview} sketches out our evaluation.
Sections~\ref{sec:naturalness},~\ref{sec:validity} and~\ref{sec:case_studies} demonstrate the naturalness of home automation, and the validity and usefulness of the scenarios generated by \tool, respectively.
Section~\ref{sec:related_work} summarizes the related work, while Section~\ref{sec:discussion} concludes with lessons learned.

%%%%%%%%%%%%%%%

%% file: motivation.tex
\section{Motivation and Background}
\label{sec:motivation}

%This paper is motivated by the need to automatically predict natural home automation scenarios, in order to address critical challenges in the design and evaluation of home automation security or safety approaches.
%In fact, 
Automatically generating natural home automation scenarios would help a variety of stakeholders
%, and not just security researchers, 
 in analyzing home automation to fulfill their own security/safety goals.
For instance, the following questions from four stakeholders could be addressed with 
%home automation 
scenarios:
\begin{itemize} 
\item {\bf Security Researcher:} What is the performance of my security system or policy, under realistic smart home usage? 
\item {\bf Platform Vendor:} Would the behavior of partner devices be compatible with my design/security policies in end-user homes?
\item {\bf Device/App Developer:} Will my security-sensitive device/app pass its unit tests in realistic user-driven automation scenarios?
\item {\bf End-user:} Can my smart home setup be unsafe in the future, based on the events that have already occurred? 
\end{itemize}

%Every question above is important for the overall health of the home automation ecosystem.
%In interest of a tractable analysis, 
We focus our motivation on one of the most important challenges in the design and evaluation of current and future systems: policy specification.
The next section illustrates how the availability of natural home automation scenarios would enhance the process of specifying home security/safety policies.

\subsection{Problem: Effective Policy Specification}
%Seamless home automation is not without its risks, as 
Prior research has built systems to discover or defend against a diverse set of security, safety, and privacy problems arising due to the misconfiguration or misuse of IoT apps~\cite{fjp16,cbs+18,cmt18,nsq+18,dh18,kmm+19}.
However, a common aspect of prior systems is their reliance on predefined security or safety policies for analysis or enforcement.
Such policies are often very intuitive, and generally created by security experts based on their understanding of the smart home.
For example, Soteria~\cite{cmt18} has a policy which states that ``the refrigerator and security system must always be on'', which is reasonable, considering the safety and security consequences, respectively, of the refrigerator or security system being OFF.
Moreover, policies created in existing literature are often carried over to future systems,~\eg Soteria's policies have been used and extended by recent systems~\cite{nsq+18,ctm19}.
That is, the effective specification of policies is significant for the design and deployment of current as well as future systems.
However, there is a gap in how policies are currently specified, as we illustrate with the following two-part motivating example:
%Consider the following motivating example, which demonstrates how a security researcher, Bob, specifies home automation security/safety policies currently, and brings out the gap in existing literature:
%More importantly, prior work has built systems that analyze routines and detect security or safety problems, at times even providing enforcement.
%For instance, Wang et al. proposed ProvThings~\cite{whbg18}, a provenance tracking system for smart homes that comes with its policy language, and is capable of enforcing certain security/safety policies.
%Celik et al. built Soteria~\cite{cmt18}, a system that analyzes IoT apps and finds safety issues in apps, prominent among which is the unintentional chaining of apps, \ie apps that may trigger other apps to accidentally execute leading to unexpected and often harmful side-effects.
%Recently, Celik et al. extended the technique proposed in Soteria to build IoTGuard~\cite{ctm19}, which enforces the security and safety policies discussed in Soteria on SmartThings apps at runtime.

%Consider the following motivating example, which demonstrates how a security researcher, Bob, specifies home automation security/safety policies currently, and brings out the gap in existing literature:

\myparagraph{Motivating Example Part 1: Manual policy specification}
Consider Alice, a security researcher who is building a system to analyze the effect of routines on user safety and security.
Alice relies on use/misuse-case requirements engineering (\eg as in prior work~\cite{cmt18,ctm19}) to come up with the policies for this system, a process that can be accomplished in three well-defined steps.
First, Alice comes up with a list of assets she cares about, \ie security-sensitive devices such as the door lock or camera, or home states that may have implications on security (\eg whether the user is home/away, if there is a fire).
Second, Alice uses her domain knowledge to come up with a set of home automation use and misuse cases, involving devices as well as environmental factors in the smart home.
This step is the hardest, as the assessment of what behavior constitutes use and misuse can often be contextual, and the possible contexts are subjective and rely on Alice's imagination. 
For instance, the camera turning OFF can be perfectly normal if the user is at home (\ie for privacy), but an example of misuse/anomalous behavior, generally, when the user is away (\ie when monitoring is needed). 
Finally, once the use and misuse cases are identified, Alice transforms them to functional requirements (\ie to ensure the use cases) or constraints (\ie to prevent the misuse cases).
However, coming up with the use and misuse cases manually is by itself a significant challenge, costing Alice a tremendous amount of time and effort.
\begin{figure}[t]
    \centering
    \includegraphics[width=2.0in]{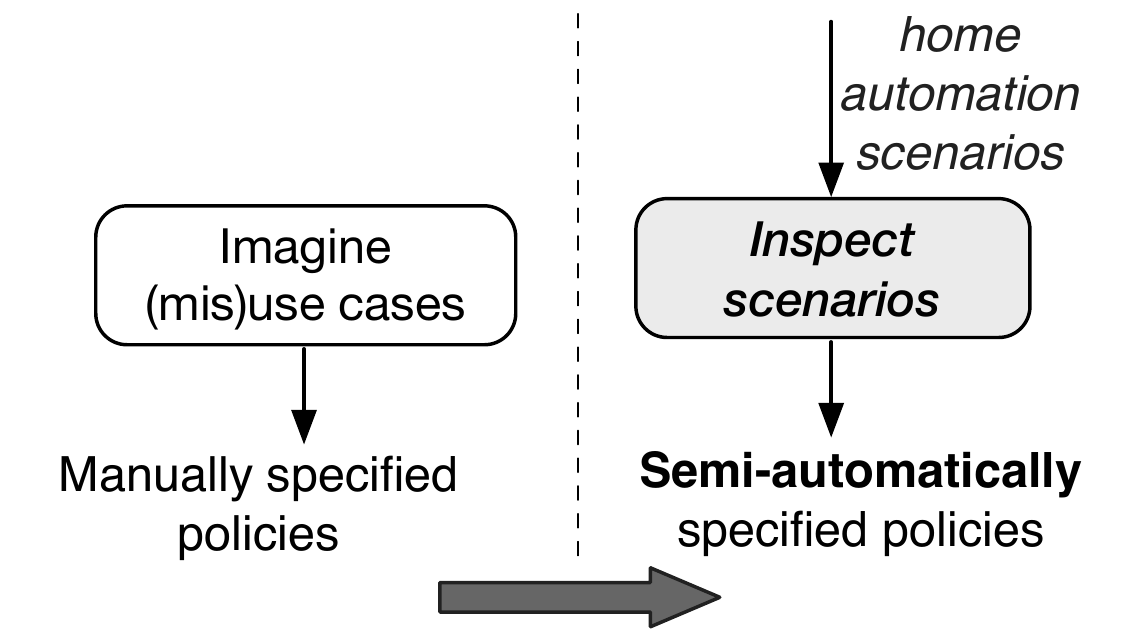}
    \vspace{-1em} 
    \caption{{\small Use/misuse case requirements engineering generally requires the analyst to imagine cases, which can be both subjective and significantly time-consuming. This process can be qualitatively improved by providing the analyst with scenarios to inspect.}}
    \vspace{-1em}
\label{fig:concept}
\end{figure}

Figure~\ref{fig:concept} shows an alternate approach that leverages scenarios, which may make Alice's task significantly easier.

\myparagraph{Motivating Example Part 2: {\em Semi-automatic} policy specification with scenarios} 
Recall that a scenario is a sequence of home automation events. 
As shown in the figure, we intend to replace the second step in requirements engineering, \ie instead of requiring Alice to {\em imagine} use/misuse cases, we assume the existence of a set of natural scenarios that are reasonably likely to occur in an end-user's home.
%we assume a framework (\ie denoted by the black box) which provides her with a set of natural scenarios that are reasonably likely to occur in an end-user's home.
Alice uses a simple state model to track the state of different assets as each event in the scenario being analyzed plays out.
Alice only {\em inspects} the state of the home when an interesting event happens (\ie a security-sensitive event such as opening/closing the door).
Inspecting policies is a simply matter of looking at a few smart home states (\ie the {\em context} in which the present event is executing), and judging whether the combination of states is safe or unsafe, with the unsafe ones resulting in policies.
The scenarios complement the existing requirements engineering process by making it semi-automated and eliminating the burden of imagining use/misuse cases.
Moreover, if the scenarios are natural, \ie reasonably likely in the wild, then the resultant policies also become much less subjective than before.

Now that Alice can use scenarios to specify policies, the key question is: {\em how can we generate natural home automation scenarios?}

%This example clearly demonstrates how natural home automation scenarios are pivotal in 
%The scenarios enabled by \tool are the pivotal aspect of this semi-automated approach, which eliminates the burden of imagining all possible use cases, and transforms it into the tractable task of inspecting certain interesting states.

% Use-misuse case requirements engineering
% Has to imagine use cases, and misuse cases. 
% what constitutes use and misuse
% 	is very contextual
% The use-misuse cases, as well as the context, are often limited by Bob's imagination.
% "what all could happen?"

%UD Routines
% How many users do yeti and yonomi have!
%Security queries
% Few of these questions can be answered by a collecting a lot of data, and simply analyzing routines. However, we need more analysis for most.
%Funny how all of these questions can be answered by predicting probable/improbable event sequences.

%% file: background.tex
%UD Routines
% How many users do yeti and yonomi have!
%Security queries
% Few of these questions can be answered by a collecting a lot of data, and simply analyzing routines. However, we need more analysis for most.
%Funny how all of these questions can be answered by predicting probable/improbable event sequences.

\subsection{Intuition: Leveraging the Naturalness in User-driven Home Automation}
\label{sec:intuition}

The objective of this paper is to develop a framework for {\em predicting} natural home automation scenarios.
In simple terms, we want to build a model that can predict sequences of events that are most likely to happen in the future, based on the events that have already happened (\ie the {\em history}).
 %in a home \DENYS{should we emphasize here HOMES rather than a home?} 
Statistical language models (LMs) enable exactly this type of prediction.
%This section provides the intuition behind using statistical language modeling for predicting scenarios, and the primitives we use in the paper.

%As described previously, 
Our core hypothesis (Sec.~\ref{sec:intro}) is that the home automation event sequences resulting from an ordered scheduling of user-driven routines are {\em natural} and exhibit patterns that may aid in prediction.
The reason behind such naturalness is intuitive as well as an extension of an argument that has revolutionized the areas of NLP and source code analysis: while languages themselves offer tremendous expressibility, {\em actual use} of languages by humans is often repetitive enough to be reasonably predictable. 
For example, in the sentence ``You only live <missing word>.'', it is easy to predict that the missing word should be ``once'', despite the fact that technically, ``barely'', ``infinitely'' or any other adverb would have been just as syntactically correct.
%Consider the following example sentence to illustrate this point: ``You only live <missing word>.''. 
%Even with a limited knowledge of pop culture, it is not hard to predict that the missing word should be ``once'', despite the fact that technically, ``barely'', ``infinitely'' or any other adverb would have been just as syntactically correct.
A language model trained on corpora of the actual ``utterances'' or use of English would be very likely to make the same prediction, simply based on the statistical probability of the word ``once'' {\em given the preceding phrase}. 

In a similar vein, home automation events generally follow predictable patterns. Imagine the following sequence of events in a room where a light is controlled by a motion sensor: ``motion sensor detects motion, lights are turned ON, motion is not detected for a while, <missing event>''. Intuitively, here the concluding event would most likely be ``lights turn OFF''.
To leverage this naturalness and make useful predictions, we use statistical language modeling.

\subsection{Background: Statistical Language Modeling and n-gram models}
\label{sec:background}
%\label{sec:routines-naturalness}

A statistical LM, fundamentally, measures the probability of a sentence, given the probabilities of the individual words in the sentence, previously estimated from a training corpus. 
That is, traditionally, a statistical LM is defined as a probability estimation over units of written/spoken language, which measures the probability of a sentence $s=w_1^m=w_{1}w_{2}...w_{m}$ based on word probabilities. 
This ability can also enable prediction, \ie by predicting the next most probable word that can follow a sequence of words. 
In the context of modeling smart home routines, we define a ``sentence'' to represent a sequence of home automation events, wherein the ``words'' (a.k.a {\em tokens}) are atomic smart home events (\eg 
%\textcolor{blue}{\sequence{<MotionSensor,motion,active>, , <MotionSensor,motion,inactive>}}
\sequence{<LightBulb,switch,ON>}).

%\TODO{@Sunil give a small example}). 
Thus, our approach will measure the probability of an event sequence $s=e_1^m=e_{1}e_{2}...e_{m}$ according to its constituent events $e$, relying on the chain rule of probability, as follows:
%\DENYS{While Eq 1, 2, and 3 make everything clear, not sure we have space}
%\begin{equation}
\label{eq:LanguageModel}
\begin{align}
\begin{split}
p(e_1^m) &= p(e_1)p(e_2|e_1)p(e_3|e^2_1)...p(e_m|e^{m-1}_1) \\
&=\prod_{i=1}^m p(e_i|e_1^{i-1})%\vspace{-0.3em}
\end{split}
\end{align}
%\end{equation}

\myparagraph{The \ngram language model}
In practice, however, there are often too many unique sentences or sequences to properly estimate the probability of tokens given long histories, even with large training corpora. 
Thus, we make use of the \textit{\ngram} language model, which assumes the {\em Markov property}, \ie instead of computing the conditional probability given an entire event or language history, we can {\em approximate} it by  considering only a few tokens from the past. 
The intuition behind n-gram language models applied to natural language is that shorter sequences of words are more likely to co-occur in training corpora, thus providing the model with more examples to condition token probabilities, enhancing its predictive power.
While \ngram models are valued for their practicality in NLP and software analysis contexts,  they are arguably an even more intuitive fit for analyzing home automation event sequences.  This is due to the organic semantic relationships smart-home events exhibit, \ie the localized {\em causal} relationship between triggers and actions (\eg a trigger preceding a corresponding action or set of actions), which are more relevant than presumably weaker correlations with events from the distant past.  
Using the \ngram model, we estimate the probability of the event sequence $s=e_1^m=e_{1}e_{2}...e_{m}$ as follows:

\begin{equation}
\label{eq:nLanguageModel}
p(e_1^m) = \prod_{i=1}^m p(e_i|e_1^{i-1}) \approx \prod_{i=1}^m p(e_i|e_{i-n+1}^{i-1})
%\vspace{-0.3em}
\end{equation}

\myparagraph{Evaluating the naturalness of a corpus} The effectiveness of \ngram modeling in the context of smart home events holds only if our intuition regarding the naturalness home automation event sequences drawn from user-directed routines is correct.
Thus, we must answer the question: \textit{Are such event sequences natural?}
Fortunately the naturalness of token sequences can be measured according to a trained model's {\em perplexity} (or its log-transformed version, {\em cross-entropy}) on unseen data. These are standard metrics used to test the viability of statistical language modeling for modeling any corpora.
A trained model will be ``perplexed'' upon observing a new sequence if it finds the sequence surprising, \ie unlike any sequence observed in the corpora. 
Thus, if a domain is natural, then the perplexity of a model built on corpora from the domain (\eg home automation event sequences from a population of users) when applied to new sequences from the same domain should be measurably low. 
That is, the model should be able to identify regularities in the event sequences, and hence, predict new sequences with confidence.
The cross-entropy $H$ of an n-gram model $M$ can be computed as follows:

\begin{equation}
\label{eq:cross-entropy}
H_M(e_1...e_n)=-\frac{1}{n}\log_2P_M(e_1...e_n)
%\vspace{-0.3em}
\end{equation}

\noindent where $H_M$ is the average negative log probability that the model $M$ assigns to each event $e_n$ in a test sequence. Perplexity is $2^{H_M}$.

%%We give an evaluation overview, this is probably not necessary:
%In Section~\ref{sec:cross-entropy-natural}, we demonstrate the naturalness of home automation event sequences by illustrating that the cross-entropy for a model trained on a corpora home automation sequences drawn from real user-driven routines is relatively low when compared to corpora from other domains.
%Section~\ref{sec:case_studies} demonstrates the usefulness of our predictions.

%% file: design.tex
\section{The \tool Framework}
\label{sec:design}

%\subsection{Design Goals}

%\subsection{The \tool Framework}

%% GENERAL OVERVIEW
\begin{figure}[t]
    \centering
    \includegraphics[width=3.3in]{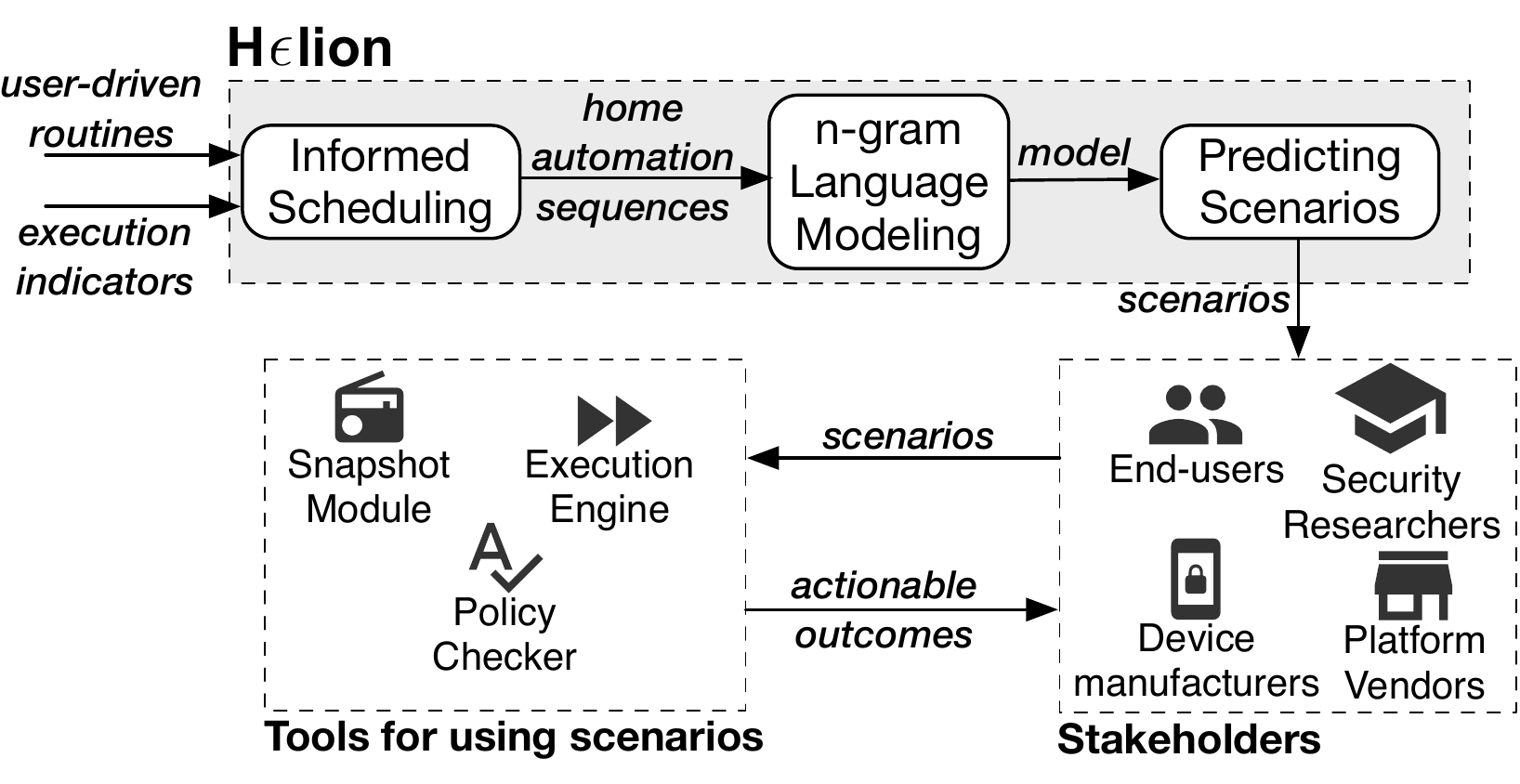} 
    \vspace{-0.75em}
    \caption{{\small An overview of the \tool framework, which models home automation sequences to construct natural scenarios. Stakeholders use tools that analyze or execute scenarios to obtain actionable outcomes.}}
	\vspace{-1.25em}
\label{fig:overview}
\end{figure}

Figure~\ref{fig:overview} shows \tool, a data-driven framework that models the regularities of user-driven home automation, generates natural home automation scenarios, and provides stakeholders with tools to use the scenarios and obtain actionable outcomes. 
As shown in the figure, the process begins by collecting {\em user-driven routines}, as well as the corresponding {\em execution indicators}, \ie clues about when or how frequently the routines may be scheduled to execute, from end-users.
\tool transforms the routines and the corresponding execution indicators into home automation {\em event sequences}, using a process called \textit{informed scheduling}.
The resulting event sequences serve as a training data set that is represented using \ngram language modeling.
\tools language model captures the patterns among the home automation sequences, and can confidently generate new natural events that follow a home's previous history. % (given separately, \ie not from the training data).
A sequence of such predicted events forms a scenario.

\tools underlying language model and sequence generation capabilities make possible a number of practical applications for improving smart home security and safety. 
We envision a set of tools -- developed to leverage \tools scenarios -- that enable these applications for stakeholders. In the scope of this paper, we present and evaluate \tools application to policy generation via a tool called the \textit{snapshot module}. This module allows security researchers to specify policies using \tool's scenarios (Sec.~\ref{sec:motivation}), by capturing the various device states of a smart home (\ie a ``snapshot'' of the home), every time a successive event in a predicted scenario is assumed to execute. We evaluate this application of \tool in depth in Sec.~\ref{sec:security_researchers}.
In addition to our main focus of policy generation, we also outline the {\em execution engine} we have prototyped, and discuss its potential use by platform vendors and device manufacturers to execute the scenarios on real devices for testing the compliance of partner devices with the their own security/safety design policies (Sec.~\ref{sec:platform_vendors}).
%The first of these tools is an {\em execution engine} developed for the SmartThings platform that enables platform vendors and device manufacturers execute the scenarios on real devices in order to test the compliance of partner devices with the their own security/safety design policies.
%Similarly, device developers can run their unit test, along with our scenarios, in the execution engine, to test the safety-related functions of their devices in reasonably real circumstances.
%We have implemented an execution engine based on the SmartThings platform, which we briefly describe in Section~\ref{sec:case_studies}.
Finally, while we do not explore this aspect, \tool could be integrated into existing {\em policy checkers} (\eg prior work such as Soteria~\cite{cmt18} or IotSAN~\cite{nsq+18}) in order to enable end-users to identify potential security/safety problems in their homes.

Next, we describe the key choices that we made while designing \tool, which is followed by our evaluation of the naturalness, validity, and usefulness of the scenarios it generates (Sec.~\ref{sec:eval_overview}$\rightarrow$\ref{sec:case_studies}).

\subsection{Collecting {\em User-driven} Routines}
\label{sec:user-driven-routines}
Recall that user-driven routines are routines created by end-users, typically with no programming experience, through interactive UIs provided by the platform (\eg the SmartThings app~\cite{smartthings}) or third-parties (\eg Yonomi~\cite{yonomiapp}).
As these routines are created by end-users, they reflect user requirements {\em directly}, relative to the marketplace IoT apps designed from a developer's perspective. \tools language modeling and sequence generation is rooted in user-driven routines. Thus, an important step in instantiating these models is the collection and representation of the events that constitute user-driven routines into semantically meaningful tokens suitable for modeling. We describe this procedure in the remainder of this subsection.

%The user-driven routines collected for our evaluation provide empirical evidence to support the above argument, \ie we discovered that out of the \xy routines collected from \xy user, only \xy or \xy\% were functionally represented in the \xy IoT apps from the SmartThings marketplace (see Section\xy for details).
%Moreover, the adoption of specific routines or combinations of routines may differ among users, which further motivates the study of user-driven routines.
%We provide additional supporting evidence in Section~\ref{sec:survey}.
%These arguments support the idea that an analysis that accounts for user-driven routines is likely to {\sf (1)} reflect realistic home automation scenarios driven by user requirements, and {\sf (2)} discover aspects of home automation that may not be visible by only examining IoT apps.

\myparagraph{1. Collecting routines from users} To collect routines from users, we use a survey methodology that is conceptually similar to prior work~\cite{ump+14}.
We thoroughly describe this methodology in Sec.~\ref{sec:data_collection}, with additional survey artifacts provided in Appendix~\ref{app:survey_methodology}.
At the end of this survey process, the raw data collected from users consists of two components: (i) routines specified in a structured natural language format, and (ii) clues for when these routines will typically execute called \textit{execution indicators}, which are used later in \tools informed scheduling process. Consider this (raw) example routine taken directly from our dataset: 

\begin{center}
	\routine{the motion is detected}{turn the light bulb on}
\end{center}

\noindent In order for smart home events to be suitably modeled by \tools n-gram language models, these structured natural language routines must be converted into semantically equivalent tokens. In a traditional statistical language model applied to natural language, the vocabulary of tokens is typically defined as the unique set of \textit{words} that exist within a corpus. 
To derive a similar set of semantically unique \textit{events} for a smart home, natural language descriptions of the same event that differ slightly (\eg\ {\sf \small motion is captured} vs. {\sf \small motion is detected}) must be resolved into the same token, following the token structure we define below. \tool automates certain aspects of this procedure to make for a semi-automated conversion (see Sec.~\ref{sec:data_collection}).

\myparagraph{2. Representing smart home events as tokens}
%As described previously, \tool uses \ngram modeling to identify the regularities in a corpus of home automation sequences composed of tokens.
In the context of this paper, the tokens are home automation events parsed from structured natural language descriptions of user-driven routines. 
% and the corpus is made up of home automation event sequences from multiple users collected. 
A home automation event can denote a change in the state of a device (\eg lock the door) or the home (\eg the user is away).
To model the varying attributes of home automation events, we express our tokens as ordered information lists called tuples. 
The design of the fields of our tokens dictates their overall uniqueness and granularity. Therefore, to effectively represent home automation events, the design of the tokens must strike a balance between encoding enough salient information to be descriptive, while still representing semantically similar events in equivalent tokens to capture meaningful patterns in the data.
% for modeling home automation. 
%We define the following format for our home automation event token:
We define \tools home automation event token as:

\begin{center}
	$e_i :=<device_i, attribute_i, action_i>$
\end{center}

\noindent where $device_i$ represents the device (\eg door lock, camera), the $attribute_i$ corresponds to one of a predefined set of device attributes (\eg the {\em lock} attribute for the door lock, which can take the values {\sc Locked}/{\sc Unlocked}), and $action_i$ represents the change of state that has led to the event. 
%and $type_i$ denotes whether the user-driven routines used this specific event as a ``trigger'' or an ``action''.
For example, an event token for the trigger event ``If the door lock is locked'' is represented by the following tuple: $<door\_lock, lock,${\sc Locked}$>$.
Note that to represent an event that causes a change to the overall state of the home (\eg ``If the user is home''), we do not use the first (\ie device) field, instead only using the attribute (\ie ``home'') and the specific change (\ie~{\sc Home} or {\sc Away}), to generate a token as follows: $<\phi,locationMode,${\sc Home}$>$.
Using this design, the example routine discussed above (\ie the motion sensor/light bulb routine) would be tokenized as follows:

\begin{center}
%\routine{the motion is detected}{turn the light bulb on} $\rightarrow$\\
\noindent {\centering $<motion\_sensor, motion,${\sc detected}$>$, $<light\_bulb, switch,${\sc ON}$>$}
\end{center}

Of the several considerations that went into the design of this token, the most important was the decision to exclude a classification of an event as a \textit{trigger} or an \textit{action}. 
This decision considers that we had to model home automation ``event sequences'' rather than a set of routines.
That is, many events can be both triggers and actions, depending on how/where they are used in a routine; \eg the trigger condition ``if the door is {\sc Locked}'', and the event resulting from the action ``{\sc Lock} the door'', deal with the same device, attribute, and device-action, but may be either a trigger or an action depending on where they are used in the routine.
This decision also prevented unnecessary sparseness in our dataset, \ie as representing the same home automation event (\ie the door being locked) using a token format that included a trigger/action specification would generate two separate tokens.
%, based on whether it is used as a trigger or an action. 
%This may lead to unnecessary sparseness in the dataset.
%Finally, choosing not to specify events as triggers or actions is also a logical choice given the task of modeling home automation ``event'' sequences, rather than a set of routines.

\input{event_sequences}
\input{modeling}

%% file: event_sequences.tex
\subsection{Generating Home Automation Event Sequences with {\em Informed} Scheduling}
\label{sec:informed_scheduling}
%\tool generates home automation event sequences from routines collected from end-users.
In order for \tool to model meaningful sequences of smart home events, the individual routines collected from users must be arranged into temporally meaningful sequences. \tool transforms the tokenized routines specified by a particular user into a home automation event sequence, \ie an approximate representation, in terms of a sequence of event tokens, of how the routines would execute in the user's home, for a given period of time. 
The ``order'' in which routines may execute in a home in any given day becomes important for the purpose of constructing this approximation. 
This importance of order can be readily observed in natural language, \ie while single words carry with them isolated meaning, a combination of words in a sentence with specific, intentional ordering form a more expressive collective meaning.
The same can be said about home automation, \ie where home automation events are like words in a sentence, and the order of such events is bound to affect the ``meaning'' (\ie implications) of the home automation sequence. 
To illustrate this point, consider a simple sequence of two events:

\begin{center}
	\sequence{motion detected $\rightarrow$ light turned ON}
\end{center} 

\noindent Intuitively, we can interpret this sequence as {\em meaning} that the light turns ON because motion is detected.
On the contrary, if the order was reversed, the sequence would not have a readily apparent logical meaning.
%In fact, if the light turned ON, and then motion was detected, we would potentially not be able to assign a meaning at all. 
This simple example demonstrates how generating event sequences and when scheduling routines in the right (or best-effort) order is important.
The question is: {\em How can we obtain this order?}

%Discuss the algorithm and scheduling indicators for generating event sequences

%Let us consider an event sequence in Bob's smart home: it is 8 a.m., blinds open, Bob leaves the house, Door is locked, smoke is detected, smoke alarm is triggered.  
%In most cases, sensors are used as triggers, which causes an event to execute on action devices e.g., Smoke detection triggering a Smoke Alarm makes more intuitive sense than a Smoke Alarm causing the smoke to be detected. 
%These orders that execute in the smart home can be clearly captured by user-driven routines where the users specify certain action based on a trigger event.

\myparagraph{1. Introduction to Execution Indicators}
We propose a novel abstraction for users to stipulate the approximate order in which routines may execute, \ie routine-specific {\em execution indicators}.
That is, we consider the possibility that end-users have some intuition regarding {\em when} certain routines execute, based on when certain device or environmental events may generally occur. 
For instance, blinds are usually opened in the morning, and closed at night.
As a result, a user may order a routine triggered by the opening of the blinds {\em before} another triggered by their closing. 
Similarly, users may be able to describe when they perform certain personal tasks which trigger home automation, \ie when they come home, go to work, bed, cook, or do laundry.
Execution indicators allow us to capture these and other such factors, which we then leverage to schedule routines to create home automation sequences.
This is why we define the approach as informed scheduling, as the {\em scheduling mechanism is \underline{informed} by the user's understanding of their own home use}.
%Note that we do not recommend tracking users for this purpose.
%Instead, 

%Note that it would be unwise to track the user's daily tasks to achieve our goal, as it would not only be prohibitively costly and privacy-invasive.
%Now, we do not recommend cataloguing the user's daily tasks to get the order in which routines may execute, as that would have prohibitive costs in terms of feasibility and privacy.
%Instead, users may be able to provide us with clues regarding this ordering by specifying {\em execution indicators} for each routine they provide, allowing us to {\em approximate} the order of routines.

\myparagraph{2. Specifying Execution Indicators and Scheduling Routines}\\
Execution indicators constitute the time and frequency of the potential execution of a routine. 
As users may not be able to specify precise values, we collect such indicators by allowing users to pick broad ranges of values organized into three types: {\sf (1)} the {\em time-range} indicator (\eg early morning, noon, and night), {\sf (2)} the {\em day-range} indicator (\eg mostly on weekdays, and mostly on weekends), and {\sf (3)} the {\em frequency} indicator (\eg many times a day, few times a day, few times a month).
As mentioned earlier execution indicators are collected from users for each routine during the data collection survey (see Sec.~\ref{sec:data_collection}). 
We then use these execution indicators to generate a month-long time-series for each user, where each routine may occupy one or more one-hour time-slots (\ie depending on frequency), using the algorithm for informed scheduling, as follows:

%\KEVIN{Maybe add algorithm pseudo-code for this?} 
%We now briefly describe our algorithm for \textit{informed scheduling} of routines using execution indicators:
We initialize a month-long time-series, with hourly slots that can hold routines.
We first place the routines triggered at specific times (\eg at 8AM, open the blinds) as defined. 
For each remaining (\ie un-placed) routine, we identify the potential slots for placement, based on its time-range indicator.
The frequency indicator of the routine determines how many instances of the routine to uniformly distribute among those slots.
This distribution is also adjusted, based on the day-range indicator, \ie skewed in favor of weekdays or weekends.
For the few routines without execution indicators (\ie when users are unsure), we randomly distribute them throughout the month in the remaining slots.
Finally, once all the routines have been scheduled in the time series, we extract the ordered set of routines from the time series as the execution sequence. In \tool, this process is automated by parsing execution indicators for each routine collected in the survey and then programmatically performing the scheduling and sequence extraction. 

We later empirically demonstrate that users can confidently supply execution indicators for most routines, only being unsure for generally unpredictable events such as fires/CO leaks (Sec.~\ref{sec:data_insights}).
Moreover, with these approximations of how routines execute, \tool can create valid (Sec.~\ref{sec:validity}) and useful (Sec.~\ref{sec:case_studies}) scenarios.
%In Section~\xy, we empirically demonstrate that users can confidently supply execution indicators for most routines, being unsure only for unpredictable events such as fires or CO leaks.
%Finally, we recognize that users may be able to indicate the schedule of some routines more confidently than other routines (as we show in Section~\ref{sec:survey}).
%%\TODO{Add reference to the validity section here.}
%and While we are able to take advantage of these approximate user-defined schedules via our execution indicators, 
{\em Accurately} scheduling all possible routines is a broader research challenge that is beyond the scope of this paper, as we discuss in Section~\ref{sec:discussion}.

%% file: modeling.tex
\subsection{Modeling Event Sequences}
\label{sec:modeling}
 
\tool uses the \ngram model to learn the regularities in user-driven home automation sequences, \ie it estimates the probability distribution of \ngrams in a corpus of the home automation sequences created previously.
For estimating probabilities, \tool follows the approach described previously in Section~\ref{sec:background}.% which describes why choosing $n\ge 3$ is reasonable for modeling home automation, as well as the need for smoothing techniques to account for data sparsity when using higher values of $n$.
 
%%%% CHOOSING N%%%%%
%\subsubsection{How to choose $n$?}
\myparagraph{Why do we need $n\ge 3$?}
Recall that when estimating the probability of a sequence of length $n$, the \ngram model computes the probability of the $n^{th}$ token appearing after the $n-1$ previous tokens (\ie the {\em history}).
The intuition behind looking back at the $n-1$ events is that they provide the {\em context} as to why the $n^{th}$ event is being scheduled.
Given this intuition, one thing is clear: when choosing $n$ for modeling home automation sequences, we can rule out values  of $n<3$.
That is, $n=1$ will only estimate the probability of individual events in the corpus, completely ignoring the context.
Choosing $n=2$ is only slightly better, as it may mostly capture simple relationships that are already observable from data, \ie trigger-action routines we collect from users.
Only with larger values of $n$, \ie $n\ge 3$, the model can learn non-obvious regularities in home automation corpora.
To illustrate this point, consider the following example sequence whose probability is being estimated using a 4-gram model, \ie $n=4$:\footnote{We use natural language instead of tokens when describing event sequences.}
%Similarly, when using the trained model, \ie predicting the next token to appear after a history, we consider the $n-1$ events in the history, and predict the $n^{th}$ token as the most probable one given the previous $n-1$ tokens.
%This is also how \tool generates its scenarios, \ie predicting one token at a time, after a particular history (\ie like a sliding window).
%Consider the following example s
%To illustrate why examining the history is valuable, consider a 4-gram model, \ie $n=4$, and the following example sequence whose probability is being estimated:\footnote{For simplifying the explanation, we use natural language instead of tokens when describing event sequences.}

\begin{center}
	\sequence{user comes home $\rightarrow$ lights switch ON $\rightarrow$ it is evening $\rightarrow$ door locks}. 
\end{center}

%\sequence{the user comes home $\rightarrow$ lights switch ON $\rightarrow$ time changes to evening $\rightarrow$ the door locks}.
Here, the factors such as the user being home, the lights being ON and the time of the day being the evening provide the context for the occurrence of the next event, \ie the locking of the door.
As a result, examining the last $three$ events certainly helps.
However, there is a caveat: considering too much of the event history (\ie a very large $n$) may actually hurt the predictive power of the model. That is, an event that occurred earlier during the day (\eg the user going to work) would likely not share any semantic relationship to the given sequence, and hence, would not really encapsulate any regularities. 
%Thus considering too much of the event history is not likely to help, and may even hurt, the predictive power of the model.
%This observation brings forth an interesting aspect of home automation: users may engage in broad home automation tasks throughout the day (\eg an ``end of the day'' task), such that each task has events that are related by the context of that task, but may not be related to the events from tasks occurring significantly before or after.
%The state of the Light before the user arrived home may not impact the sequence of event that occurred.
As a result, the choice of $n$ directly impacts the ability of the model to capture the existing relationships between events in the corpus, especially if they are related by the virtue of belonging to the same, high-level user-activity (\eg end-of-day events).

\myparagraph{The need for smoothing} 
Selecting $n\ge 3$ may seem to intuitively lead to a better model, however, for higher orders of $n$ there will inherently be fewer sequences for the model to learn from.
This is because longer sequences tend to be unique, \ie the sequence: 

\begin{center}
	\sequence{the user leaves the home $\rightarrow$ the door locks}
\end{center}

\noindent may occur often in a set of sequences, whereas the sequence:
\begin{center}
	\sequence{user leaves the home $\rightarrow$ door locks $\rightarrow$  motion is detected}
\end{center}

\noindent is likely to be less common.
%but in practice, larger values of $n$ will cause data to become sparse, and hence, make prediction difficult. 
%a simplistic approach that predicts the next event by directly checking the probability of the given history will not work, due to the sparsity of data.
This leads to a data sparsity problem, wherein it is likely that a history queried during prediction may have not been observed in a training corpus.
As a result, a naive model will be unable to predict the next event, \ie if this entire history is not present in the training corpus, the model will not be able to make a good prediction, even if it may have observed subsequences of this history (\eg user leaving \& door locking). %, given that the history itself seems to be improbable.
%Consider a 4-gram model attempting to predict the next token from the sequence with the motion detector event above. %the following example history, \ie input/query, to a 4-gram model for prediction:
%\sequence{<the user goes away> <the door locks> <the motion sensor senses motion>}
%If the trigram above is present in the training data, then the model simply predicts the most probable event after the trigram, as estimated from the training data.
%However, if this entire history is not present in the training corpus, the model will not be able to make a good prediction, even if it may have observed subsequences of this history (\eg user leaving \& door locking). 
%Now, the model may have observed subsequences of this history (\eg the user goes away and then the door is locked) during training; however, if this entire history/input is not present in the training corpus, the model will not be able to make a good prediction. 
%In other words, if the dataset is sparse, a 4-gram LM will frequently encounter improbable histories, leading the model to be {\em infinitely perplexed}, \ie demonstrate high perplexity for most queries.
%As we increase the \ngram order, the number of possible sequences (\ie permutations of event tokens) increases as well.
%Hence, the model will have a higher chance of being queried with a sequence that has never been observed in the training set, as the training set becomes sparse as the $ngram$ order grows. 
To allow the LM to assign probabilities (\ie useful predictions) to previously un-observed sequences with sufficient statistical rigor, we rely on {\em smoothing}~\cite{chen-goodman}. Smoothing is a well-known technique in NLP where the model assigns some probability distribution to rare or unobserved sequences.
We consider two smoothing methods to improve the prediction quality of the model with higher-order \ngrams, \ie~{\sf (1)} {\em backoff} and {\sf (2)} {\em interpolation}. At a high level, backoff smoothing techniques simply revert to predictions based on lower order n-grams when an observed history of higher order n-grams are rare or haven't been observed. Conversely, interpolation always considers and combines token probabilities for lower-order n-grams when making predictions. In our instantiation of \tool, we elected to use interpolated n-grams due to their demonstrated ability to perform well with lower-order (\ie 3-4 gram) models \cite{chen-goodman}.

\input{flavors}

%% file: flavors.tex
\subsection{Generating Different ``Flavors'' of Scenarios for Security/Safety Applications}
\label{sec:predicting_test_cases}
%Security use cases
\tool generates scenarios by treating the model as a {\em sequence generator} that can produce an arbitrarily long series of smart home events, given a history.
That is, given a history, the \ngram model looks at the previous $n-1$ events, and predicts the next most probable event.
For the next prediction, the newly predicted event now becomes a part of the history (\ie the latest event in the history). 
We can continue these predictions to get arbitrarily long scenarios.
These predictions can be used to generate scenarios that are natural, \ie reasonably likely to happen in some user's home, and very likely with respect to the training data.
However, in applying \tool in practice (\eg for policy specification) it may be desirable to generate both highly natural and \textit{un}natural event scenarios.
%However, the question is, is predicting this one type of scenario sufficient for enabling security and safety-related analysis?

Consider the policy specification example from Sec.~\ref{sec:motivation}, where
Alice, needs scenarios that can replace her need to come up with use {\em and misuse cases}.
That is, Alice not only needs likely scenarios, but she also needs highly unlikely scenarios, which may not be normally observed or expected, and hence, may actually demonstrate stress tests, or rare but unsafe situations.
Prior work demonstrated that LMs can be configured to generate different {\em flavors} of event sequences~\cite{Linares-Vasquez2015}, aside from the standard natural event sequences they predict.
In a similar vein, \tool can be configured to generate two {\em flavors} of scenarios, motivated by our policy specification use case:
%In this section, we describe how we can support a variety of security evaluation scenarios by generating test cases of different flavors, \ie by making strategic changes to the model's probability estimation.

\myparagraph{The {\em up} flavor, for {\em non-adversarial} use} This flavor is the default, \ie where our model generates highly probable event sequences, given a history using the natural probability distribution over tokens in the training corpus. 
The {\em up} flavor demonstrates the ``normal'' patterns in user-driven home automation, which may be used for many evaluation tasks that require producing normal home automation scenarios.
For instance, the {\em up} scenarios may be analyzed to diagnose {\em configuration-related} safety issues in user-driven home automation.

\myparagraph{The {\em down} flavor, for {\em deliberate} misuse} The {\em down} flavor corresponds to an {\em unnatural} distribution over tokens, \ie\ {\em down} scenarios will exhibit patterns that are highly improbable given the model's probability estimation. 
\tools language model generates {\em down} scenarios by sorting the model's most probable token predictions given some history, and then reversing this order, such that the \textit{most improbable} token is given by the model as the prediction.
Thus, the {\em down} flavor can be thought of as a stress test, where highly unlikely events are purposefully predicted. 
The {\em down} scenarios may be interpreted as a system under constant attack, or where all devices are simultaneously malfunctioning, or exhibiting ``abnormal'' behavior.
%We expect the {\em down} flavor to help us generate misuse scenarios, for Alice to come up with effective security policies.

\begin{comment}
\myparagraph{The {\em Intermittent} flavor, for {\em practical} use} The {\em intermittent} flavor generally generates natural events, but, after a series of natural events have been generated, it may also predict an unnatural event.
This flavor can enable evaluation of systems under practical levels of stress or compromise, \ie where at most one or two devices may be compromised.
Moreover, {\em intermittent} test cases may allow us to analyze the behavior of the home automation (or a device) after a security/fault-related incidence has taken place (\ie after the abnormal event in the test case).
\end{comment}

%We demonstrate the usefulness of these flavors through our semi-automated discovery of security policies in Section~\ref{sec:case_studies}.

%\SUNIL{One thing I haven't talked about is how we have used multiple devices in the same token and also the "high/low/fixed\_number" that we used instead of numbers that users provided \eg ~$<Air\_Conditioner-Door\_Lock,temperature-lock,high-lock>$ }

%% file: data_collection_insights.tex
\section{Data Collection and Initial Findings}
\label{sec:data_insights}

%We perform evaluation with a training dataset constructed by obtaining routines and execution indicators from real users.
In this section, we describe our data collection methodology, our approach for constructing the {\sf HOME} corpus consisting of home automation sequences for use with \tool, and most importantly, important initial findings that can be directly gleaned from the data. 

\myparagraph{\underline{Data Availability and IRB Approval}} We plan to release anonymized datasets (\ie including the {\sf HOME} corpus) upon publication. 
Further, all the user studies and surveys described in this paper were approved by our institutional IRB.

%% Data Collection and statistics
\subsection{Methodology for Collecting Data from Users}
\label{sec:data_collection}

We use a survey methodology for collecting data from end-users.
We surveyed 40 end-users who were generally from the Computer Science (CS)
academic population: 37 participants were current graduate and undergraduate students, and 3 had PhDs. 
%Most participants (\xy) marked CS as their field of expertise, followed by software engineering (\xy), and the remaining few specified research areas within CS.
A majority owned at least one smart home device (24 or 60\%), while a significant minority had experience creating routines (17 or 42.5\%).
The {\sf HOME} corpus was constructed using the data from these users.

To make the task of specifying routines and indicators easier, we split the survey into a series of logical steps that we describe in this section.
%In the interest of space, we have made the images of the survey instrument available in the Appendix~\ref{app:survey_methodology}.
We have made the survey available in the Appendix~\ref{app:survey_methodology}.

\myparagraph{1. Selecting devices} First, participants selected devices that they could envision (or already have) in their smart home. To enable this step, we provided the participants with a broad {\em device list} consisting of 70 unique types of devices available in the market.
We constructed this list using resources such as websites and mobile apps of all the device partners of the popular Samsung SmartThings~\cite{smartthings} and Google NEST~\cite{nest}, popular technology websites, and technology forums. Fig.~\ref{fig:device-selection} in Appendix~\ref{app:survey_methodology} shows our device selection screen.

\myparagraph{2. Creating routines} After selecting devices, the participants were given a short tutorial on routines, and asked to create one or more routines using the devices that they had previously selected, along with general smart home variables such as the user being home/away, temperature, and time. We asked the participants to provide triggers and actions in a plain English text to allow them to express any functionality desired. We provided interactive form with two text boxes (i.e., for triggers and actions) as shown in Fig.~\ref{fig:routines_creation} of Appendix~\ref{app:survey_methodology}.

We provided participants with the functional information about devices in the survey, to help them focus on the task of creating routines.
As shown in Fig.~\ref{fig:attribute}, % in Appendix~\ref{app:survey_methodology}, 
participants could view the functional attributes (\eg the ``lock'' attribute for the door lock device) that were applicable to every device they had previously selected, as well as the general smart home variables.
To enable this approach, we created a {\em device-attribute map} by systematically assigning one or more of the 110 attributes that we obtained from existing platforms (\eg NEST~\cite{nest} and SmartThings~\cite{smartthings}) to each of our 70 devices.

\myparagraph{3. Specifying Execution Indicators} After creating routines, participants specified the time-range, day-range and frequency indicators for the routines they created, shown in Fig.~\ref{fig:day},~\ref{fig:week} and~\ref{fig:frequency} respectively. % in Appendix~\ref{app:survey_methodology}. 
Participants could select from predetermined ranges, as well as indicate ``anytime'' for routines that could occur at any time (\ie with respect to the time-range and day-range indicators), or ``not sure'' if they were unable to specify.
In addition to collecting routines and execution indicators, we also collected  information that may assist in understanding user-driven home automation (see Appendix~\ref{app:survey_methodology}).% (the survey instrument for these additional questions is provided in Appendix~\ref{app:survey_methodology})\SUNIL{confirm this}.

\subsection{Constructing the {\sf HOME} Corpus}

The routines created by the participants in plain English were transformed into an intermediate trigger-action format, and then tokens, using the syntax described in Sec.~\ref{sec:user-driven-routines}. 
We also considered two additional situations when tokenizing: (1) if the trigger/action consisted of a conjunction of events, we combined the events into a single token (in the alphabetical order by device), as those events would be expected to execute simultaneously, and (2) for attributes with continuous values (\eg temperature), we abstracted the user-provided values into ranges (\eg low, medium, and high temperature), to create semantically unique tokens. Two authors independently verified the correctness of the tokens.
Finally, we constructed a {\em month-long} home automation event sequence for each participant using informed scheduling 
(Sec.~\ref{sec:informed_scheduling}), creating the {\sf HOME} corpus.
%, which we describe in the next section.

\begin{comment}
\myparagraph{Sequence generation} We constructed per-user event sequences using informed execution,
%used the execution indicators to construct per-user event sequences, 
 and further segmented these event sequences into smaller ``sentences'' (\ie the finite sequences estimated by the model) of 20 tokens each, to form the home automation (\ie the {\sf HOME}) corpus.
Segmentation is a necessary and common step in NLP, but because user-driven routines do not have any syntactic indicators for sentence endings (\ie unlike software, where prior work~\cite{Hindle:ICSE12} uses a line of code as a sentence),  we choose 20 experimentally as an average-case sequence length (see Appendix~\ref{app:sequence_length}).
\end{comment}

\subsection{Initial Findings}
\label{sec:initial_findings} 

The {\sf HOME} corpus consists of 30,518 home automation events, from 40 month-long sequences (\ie for 40 users), generated from 273 routines (233 unique) and their execution indicators specified by users.
We use the {\sf HOME} corpus for our evaluation in Sec.~\ref{sec:eval_overview}$\rightarrow$~\ref{sec:case_studies}.
%Moreover, we also plan to release the anonymized data behind the statistics that lead to our findings.
%In the rest of this section, 
Next, we describe observations and findings from our data analysis.

\finding{1}{Routines are {\em important} to users. Moreover, users leverage most available devices for creating routines} 
When asked how important routines were to them, 4 users or 10\% indicated that routines were ``very important'', 20 users or 50\% indicated ``important'', while 16 or 40\% indicated ``somewhat important''.
No user indicated that routines were unimportant.

%\finding{2}{Users may use most available devices for automation} 
Further, our participants used 61 out of the 70 devices provided (or 87.14\%) in at least one routine.
%, a total of 61 (or 87.14\%) were used in at least one routine by our participants. 
Devices related to lighting and temperature control were the most popular for automation (\ie selected by 23 and 21 users respectively), closely followed by security devices such as cameras.
This finding indicates a strong user preference for creating routines and integrating device functionality.

\finding{2}{SmartApps do not represent a significant number of user-driven routines}
% Q: How many SmartApps are not in common with our set:
%% 149/187 are not in common (i.e., 38 in common)
Out of the 233 unique routines created by our 40 end-users, only 134 or 57.51\% could be represented by SmartApps from the SmartThings market~\cite{smartapp-repo}, \ie 42.49\% were not.
Moreover, only 40 out of the 187 SmartApps or 21.39\% actually accounted for the represented routines, \ie our users would potentially not use the remaining 147 or 78.6\% SmartApps.  

\finding{3}{Users indicate a strong preference for controlling/creating their own routines} 
Most users (\ie\ 32 or 80\%) indicated that their source of ideas for creating routines were {\em personal requirements}. 
Moreover, when asked about their preference for vendor-controlled versus user-controlled home automation, most users (20 or 50\%) said they would prefer a combination of both, a significant minority (\ie 17 or 42.5\%) said they would prefer a purely user-controlled home automation setup where users define routines, whereas a negligible number said they would prefer a vendor-controlled setup (3 or 7.5\%).
 
\finding{4}{Users may perceive any device as security sensitive, depending on the context} We asked participants to select devices (out of 70) they considered to lead to harm if compromised or malfunctioning (\ie are security/safety-sensitive). 
As expected, the devices directly aligned with security/safety were primarily selected, such as the security alarm (36 participants), door lock (35), camera (35) or the garage door opener (31).
Even more surprising was that every single device was marked as security/safety-sensitive by at least 3 participants; \ie participants also considered tangential scenarios where their well-being may depend on the device, even if security/safety is not its primary feature.
%Our findings (\ie\ \fnumber{1}$\rightarrow$\fnumber{4}) strongly support the need for examining user-driven routines.

\input{tables/execution_indicators}
\finding{5}{Users can confidently specify {\em certain execution indicators}, aside from certain unpredictable triggers}
Table~\ref{tbl:indicators} summarizes the options chosen by our participants, \ie selected one of the specific ranges offered, or anytime, or ``not sure'', for each of the three execution indicators (\ie time-range, frequency, day-range).
Our data shows that the participants could specify execution indicators confidently for most of their routines, \ie they were ``Not sure'' in very few cases (\ie\ {\em at most 10.26\%, for the day-range indicator}).
Moreover, users are generally able to select specific time-range and frequency values for a majority of their routines, \ie for time-range (\ie 56.78\%) and frequency (\ie 93.77\%).
This clearly demonstrates that users are able to confidently supply execution indicators for most of their routines.
On further analyzing the ``Not sure'' cases, we discovered that most are caused by triggers that are unpredictable by nature, \eg CO leaks or drastic temperature changes, which explains why users could not specify them. 

The analysis of the survey data exposes the evident need to analyze user-driven routines (\fnumber{1}$\rightarrow$\fnumber{3}), and additional challenges, such as unpredictable execution indicators (\fnumber{5}).
Sections~\ref{sec:eval_overview}$\rightarrow$\ref{sec:case_studies} will explore this data further in terms of the naturalness of the home corpus, as well as the validity, and usefulness of \tools scenarios.

%% file: tables/execution_indicators.tex
\begin{table}[t]
\centering
\footnotesize
\caption{{\small Selection of specific ranges, any time, or not sure options for
execution indicators, by number of routines.}}
\vspace{-1.5em}
\label{tbl:indicators}
\begin{tabular}{r|c|c|c}
  \Xhline{2\arrayrulewidth}
%  \rowcolor[HTML]{C0C0C0}
 {\bf Execution Indicator}	& {\bf Specific Range}	& {\bf Anytime} & {\bf Not Sure}\\ 
  \Xhline{2\arrayrulewidth}
  Time-range		& 155 (56.78\%)		& 107 (39.2\%)	&  11 (4.02\%) \\
  Frequency		& 256 (93.77\%)		& 0		& 17 (6.23\%)	\\
  Day-range		& 59  (21.61\%)		& 186 (68.13\%)	& 28 (10.26\%) \\
  \Xhline{2\arrayrulewidth}
  \end{tabular}
\vspace{-2em} 
\end{table}

%% file: naturalness.tex
\section{Research Questions (RQs)}
\label{sec:eval_overview}

The three major RQs address the core contributions of this paper: 
\begin{itemize}
	\item {\bf RQ$_1$} : How {\em natural} is home automation corpora?
	\item {\bf RQ$_2$} : Do the scenarios generated seem {\em valid} to the end-user? 
	\item {\bf RQ$_3$} : Can \tools natural scenarios be applied in {\em useful} ways to improve the security and safety of home automation?
\end{itemize}

To answer \resq{1}, we test our {\em naturalness hypothesis} for our {\sf HOME} corpus, which is the foundation of \tool (Sec.~\ref{sec:naturalness}). 
\resq{2} deals with the perceived validity of the scenarios/event sequences \tool generates. To answer this question, we perform two experiments with {\em external evaluators}, \ie 16 users who were not a part of the {\sf HOME} corpus (Sec.~\ref{sec:validity}).
%That is, \resq{1} and \resq{2} form the {\em intrinsic} evaluation of \tool. 
To answer \resq{3}, we revisit the policy specification problem from the motivating example, and demonstrate how \tools scenarios enable an expert to predict security and safety policies with manageable effort (Sec.~\ref{sec:case_studies}).
We also explore how stakeholders may gain insight into the security/safety problems in platforms/devices under realistic circumstances, by executing the scenarios in an execution engine based on the SmartThings platform.

\section{Evaluating the Naturalness of {\sf HOME} (\resq{1})}
\label{sec:naturalness}

We test our naturalness hypothesis by comparing the cross-entropy of real user-driven home automation sequences with that of a popular natural language and software corpora. 
Recall that cross-entropy is a measure of how perplexed a model is when it is exposed to a sequence from a corpus of the same domain from which the model was trained (Sec.~\ref{sec:background}).
We measure cross-entropy of the {\sf HOME} corpus, using the MITLM toolkit~\cite{mitlm}.
%\footnote{We plot the values for interpolation smoothing, as backoff smoothing shows values that are similar, \ie less than 2 orders of magnitude different.}
To properly measure the cross-entropy of our corpus, 
%we must hold out some of the data for testing, while using the remainder to train the language model. Hence, 
we perform 10-fold cross-validation. %, \ie randomly split the corpus into 90\%/10\% sequences, train (\ie estimate probabilities) on 90\% of the data, and test (\ie predict probabilities) for the 10\%. 
%\KEVIN{We may want to move our explanation of Cross Entropy from Section 2 to here. I think it might flow better.} 
Moreover, since naturalness is relative, we compare the cross-entropy of the {\sf HOME} corpus for different n-grams with that of the Gutenberg corpus (\ie English), as well as a software~(C\#~\footnote{The C\# corpus shows the best cross entropy out of all the programming corpora~\cite{Hindle:ICSE12}}) corpus by Hindle et al.~\cite{Hindle:ICSE12}.\footnote{We obtained the raw cross entropy values for the C\# and Gutenberg corpora from ~\cite{Rahman:ICSE'19}}

\begin{figure}[t]
    \centering
    \includegraphics[width=2.2in]{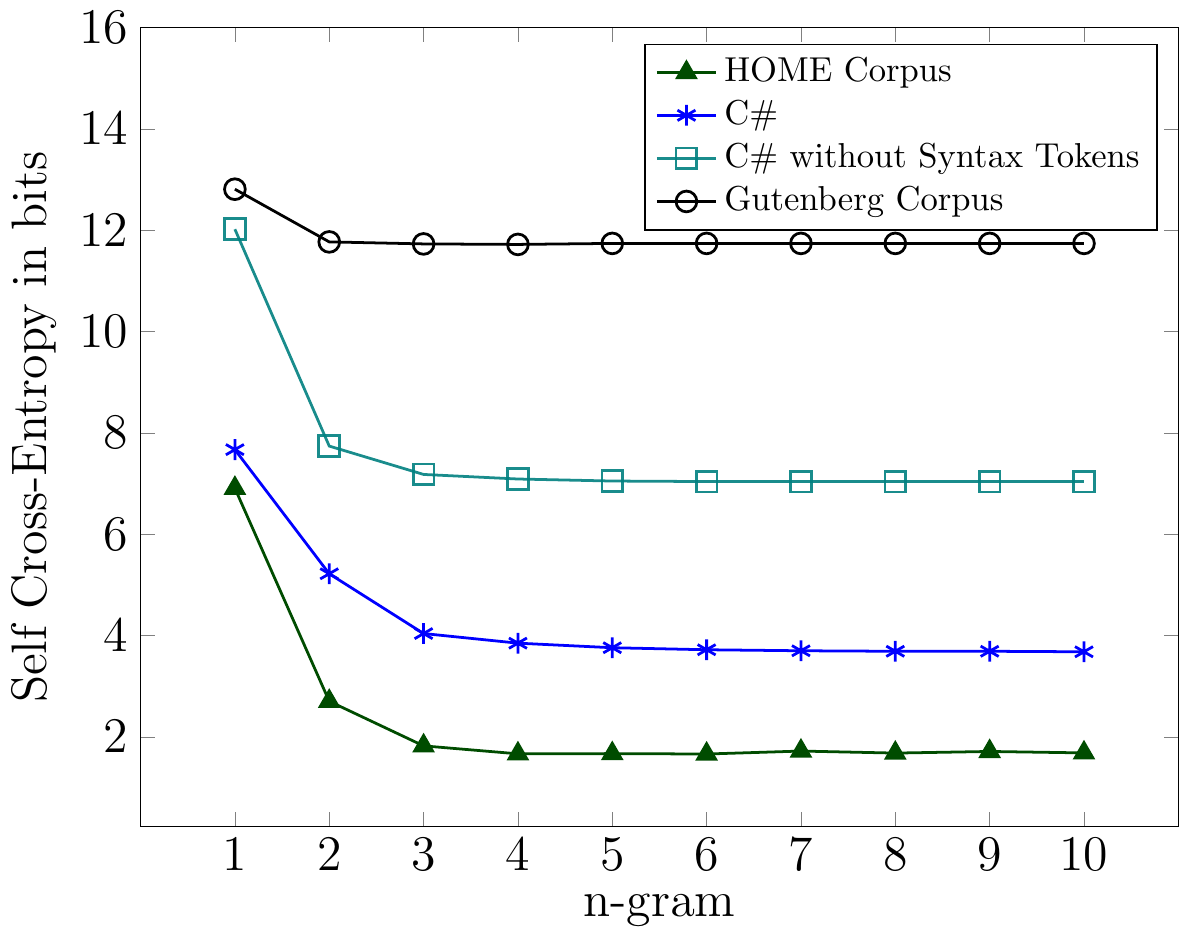}
    \vspace{-1.25em} 
    \caption{{\small Cross-entropy of the \ngram model over the {\sf HOME}, Gutenberg, and C\# corpora, as well as the C\# corpus {\em without syntactic tokens}.}}
	\vspace{-1.5em}
\label{fig:plots_all}
\end{figure}

\myparagraph{Results} As seen in Figure~\ref{fig:plots_all}, the cross-entropy for the {\sf HOME} corpus starts at a high of 6.9 bits for a unigram model, drops to 2.70 bits for the bigram model, and then stays close to 1.7 bits for the rest of the values of $n$ from 3 up to 10.
There is an explanation for this trend: As our corpus is generated from trigger-action pairs (\ie user-driven routines), a unigram entropy is expected to be much higher, as the unique token frequencies may be heavily skewed.
However, as the model considers more history (\eg for the bigram), the entropy drastically reduces. 
This trend is observed in the Gutenberg and C\# corpora as well.
We now discuss the key finding from our analysis:
\begin{comment}

\begin{figure}[t]
    \centering
    \includegraphics[width=2.3in]{figs/plots_with_syntax} 
    \vspace{-1em}
    \caption{{\small Cross-entropy of the \ngram model over the {\sf HOME}, Gutenberg, and C\# corpora.}}
    \vspace{-1em}
\label{fig:interpolation}
\end{figure} 

%\footnote{We obtained these values from \xy et al.~\REF}, using interpolation smoothing.   
\begin{figure}[t]
    \centering
    \includegraphics[width=2.3in]{figs/plots_without_syntax}
    \vspace{-1em} 
    \caption{{\small Cross-entropy of the \ngram model over the {\sf HOME}, Gutenberg, and C\# corpora {\em without syntactic tokens}.}}
	\vspace{-1em}
\label{fig:updated_hindle}
\end{figure}
\end{comment}

\finding{6}{The {\sf HOME} corpus is natural relative to English language and software corpora, {\em without any syntactic glue}}
As seen in Figure~\ref{fig:plots_all}, the {\sf HOME} corpus has much lower cross-entropy values, and can be said to be more natural, which satisfies our {\em naturalness hypothesis} (\resq{1}).
Further, there is another interesting aspect of the {\sf HOME} corpus: the user-driven home automation sequences in the {\sf HOME} corpus are purely semantic in nature, \ie have zero syntax involved, and only capture the functionality that the user desires from the automation. 
On the contrary, C\# or any other software corpora sometimes appear more natural than they are because of the common ``syntactic glue''~\cite{Rahman:ICSE'19}, which leads to a steep drop in naturalness if the glue is removed (see ``C\# without Syntax Tokens'' in Figure~\ref{fig:plots_all}).
That is, while the naturalness of software corpora will drop when the syntactic glue is removed, the {\sf HOME} corpus will be unaffected, which indicates it's amenability to being predicted.

This section demonstrates that our model captures the regularities in the {\sf HOME} corpus.
Thus, the scenarios generated by \tool will be natural (\ie likely based on the observed sequences).
However, {\em are such scenarios {\em valid} according to end-users}?
The next section addresses this question with two studies with external evaluators.
%Moreover, that home automation sequences are more natural than software is even more encouraging, and could motivate interesting use cases as previously seen in the domain of software engineering.
%While we do not claim generality (\ie as the {\sf HOME} corpus may or may not reflect the general preferences), the next two sections demonstrate how a model trained on the {\sf HOME} corpus can be effectively leveraged to produce valid routines from the users' perspective, and to help stakeholders.
%how the modeling software led to improvements in tasks like code completion, and that we show the advantages of this naturalness applied to real security tasks in the next section.

% dataset specs
	% how many users
	% how many responses considered
	% how many routines
	% experience of users: routines, smart homes, highest education, expertise
	% we further explore 

% routines were created in natural language --> manual labeling of the data into tokens
% automated transformation into ha sequences as described in Section~\ref{sec:design}

% cross-validation

% baseline (english dataset obtained from prior work)

%% Findings

%% Survey Results

%% file: validity.tex
\section{Evaluating the Validity of the Scenarios (\resq{2})}
\label{sec:validity}

%Our previous experiment demonstrates that home automation corpora are indeed highly natural (\ie in the sense of having patterns), and thus amenable to being effectively modeled using n-gram language models.
As the intended use case for \tool is the generation of scenarios for a variety of security applications, it is equally important to understand how {\em valid} (\ie reasonable) the scenarios generated by \tool are in the view of end-users (\resq{2}).

We assessed the ability of \tool to generate valid scenarios with 16 {\em evaluators}, \ie a separate set of users who were not included in the initial 40 users described in Sec.~\ref{sec:data_collection}, but from the same user population. 
This evaluation consists of two studies.
First, we conduct the {\em routine comparison study} (Sec.~\ref{sec:routine_comparison}), \ie we evaluate whether \tool can generate routines that are just as valid as the routines collected from users, in the view of the evaluators. 
This allows us to reason about validity in comparison with a real baseline.
We then move on to evaluating the sequences generated by \tool, \ie the scenarios, with the {\em sequence generation study} (Sec.~\ref{sec:sequence_generation}).
That is, we generate scenarios using histories given by the external evaluators, and test their validity under various model configurations.  
The survey instrument used in these experiments is illustrated in Appendix~\ref{app:user_studies}.
%We now describe these studies and our findings.
%Can \tool generate routines perceived to be as natural as {\em real} user-driven routines? 
%: (1) we sample sequences of size two from \tools scenarios (\ie essentially, routines), and evaluate whether users find them valid, relative to routines created by real users, and (2) we generate scenarios using histories given by the external evaluators, and test their validity, one event at a time (Sec.~\ref{sec:validity})
%These users are from the same population (\ie academic, mostly CS students), as the {\sf HOME}.
%We performed two user studies, \ie the {\sf (1)} {\em routine comparison} study (Sec.~\ref{sec:routine_comparison}) , and the {\sf (2)} {\em sequence generation} study (Sec.~\ref{sec:sequence_generation}), which helped uncover distinct aspects of the validity of \tools scenarios, as well as uncovered new information regarding how users perceive home automation.
%The full dataset associated with these studies, including all generated routines and anonymized participant responses will be made available upon publication. 
%
%\SUNIL{We need to add the survey instrument for these questions to the appendix, and add references here.}

\subsection{Routine Comparison Study}
\label{sec:routine_comparison}
We envision that many applications of \tool will rely on its ability to generate reasonable bigrams, \ie pairs of events which may be analyzed as routines.
%As a result, bigrams generated using \tool must be generally reasonable.
Using the routines previously created by end-users as our baseline (Sec.~\ref{sec:data_collection}), this study answers the following key question: {\em does \tool generate routines that are as valid as routines created by real end-users?}

\myparagraph{Generating Routines} 
We used the {\sf HOME} corpus to generate routines.
Recall that \tool is a sequence generator, and will predict a sequence of events, one event at a time, if given a certain history (Sec.~\ref{sec:predicting_test_cases}).
We predict routines using an approach similar to 10-fold cross validation: \ie in every round, we split the dataset randomly into 90\%/10\% sequences, train on the 90\%, sample histories from the 10\% testing sequences (\ie pick random subsequences of ``odd'' length), and use them to generate sequences.
Since these histories are of odd lengths, it is likely that the first prediction using any such history will be the completion (\ie the action) of a routine, the trigger of which was the last event in that history.
As a result, we assume the second and third generated events as a fresh routine. %, which we consider for analysis.

We randomly generated 40 such unique routines, generated through the following configurations of the model (\ie varying the \ngram and {\em up}/{\em down} modalities): {\sf (i)} {\em up/3-gram} (10 routines), {\sf (ii)} {\em up/4-gram} (10 routines), {\sf (iii)} {\em down/3-gram} (10 routines), and {\sf (iv)} {\em down/4-gram} (10 routines). 
We chose the {\em up} flavor to answer the validity question, but we also chose the {\em down} flavor to understand if really unnatural routines (\ie down) are also perceived by users as invalid.
%We then randomly pick unique sequences from this generated set, and sample the second and third predicted events as a routine.
%To generate routines, we use the 10-fold cross-validation approach (as we did in Section~\ref{sec:naturalness}), but provide the random sequences taken out for cross validation as histories to \tool, and generate new sequences.
%a history of length 5, which is drawn (\ie and hence removed) from the training corpus, and generate a sequence of length 3, {\em selecting the last two generated events as the routine}. 
%Selecting the second and third generated event allows us to select a fresh routine, \ie selecting the first generated event result in simply finishing a routine that started at the end of the history, which is why we select the next two.
%Using the above approach, we generated 100s of routines, and collected a random sample of 40 routines, in the following configurations (\ie varying the \ngram and UP/DOWN modalities):
Additionally, we randomly selected 10 unique routines from 10 different users from the data collected in Sec.~\ref{sec:data_collection}, \ie\ {\sf (v)} {\em survey} (10 routines).

\myparagraph{Methodology} We conducted this study in-person, wherein a proctor explained the study procedure to the evaluators, and noted down any comments. 
Evaluators were asked to rate each of the 50 unique routines described previously, in terms of how valid (\ie reasonable) they seemed, according to a modified Likert Scale (\ie with options: Strongly Agree, Somewhat Agree, Somewhat Disagree, and Strongly Disagree). 
The order of routines was randomly shuffled for each evaluator to mitigate inductive bias.
%, such that no two evaluators received the routines in the same order, 
% between users, and the study was structured in such a manner that participants were evenly exposed to each type of routine first, thus mitigating inductive bias.
Additionally, evaluators were given the opportunity to provide additional feedback after the task.
%, which resulted in some of the interesting findings.
%Given that the definition of what a ``natural'' routine means could vary from participant to participant, the study proctor provided an explanation with clearly natural, and clearly unnatural examples to effectively illustrate the concepts to participants.

\input{tables/validity-study}

%\myparagraph{Findings} 
\myparagraph{Results} Table~\ref{tab:bigram-study-results} summarizes the ratings for the 10 routines in each configuration by 16 evaluators (\ie 160 ratings per configuration). The evaluators generally rated the routines generated using the {\em up} flavor as well as the routines from the survey as valid/reasonable, whereas the routines generated using the {\em down} flavor were mostly rated as invalid. We now describe our findings:
 %in each configuration (\ie ratings 16 evaluators)), in terms of the percentage of users that supported that rating. 
%We now discuss the findings from this table, as well as the additional comments supplied by our participants.

\finding{7}{Routines generated using \tool are as valid as routines created by users}
%Observe the UP (3 and 4-gram) ratings, as well as the ratings for the routines from real users (\ie SURVEY), in Table~\ref{tab:sequence-study-results}.
Our evaluators rate over half (\ie 50.63\%) of the {\em up/3-gram} routines and 69.38\% of the {\em up/4-gram} routines as strongly agree. In the latter case, \tool actually scores higher than the 63.75\% strong agreement for the {\em survey} routines.
%which is comparable to the 63.75\% strong agreement with the {\em survey} routines (and higher in case of {\em up/4-gram}).
Moreover, the total agreement (\ie strongly plus somewhat agree) is just over 70\% for {\em up/3-gram} routines, and 82.5\% for the {\em up/4-gram} routines, the latter of which reasonably valid, relative to the 88.75\% agreement for the {\em survey} routines.
%These results indicate that \tools routines, especially for the 4-gram model, are perceived as reasonably valid relative to {\em survey} routines. 

\finding{8}{{\em down} routines are generally perceived as invalid, and sometimes perceived as unsafe/insecure}
Both {\em down/3-gram} and {\em down/4-gram} routines are overwhelmingly rated as invalid, with 68.13\% strong disagreement in both cases. 
This implies that the unnatural sequences generated by \tool are also generally invalid according to users.
This finding for routines is likely to extend to longer scenarios as well.
More importantly, some users commented that these routines were unsafe; \eg one user said that ``{\em ...automations are a hazard. If I'm not home I don't want to gas stove to turn on.}''
This finding validates our approach of using highly unnatural sequences for stress testing or modeling adversarial misuse cases. 

\finding{9}{The notion of validity may vary due to personal preferences, even among users from the same population} Evaluators rated 7.5\% of the {\em survey} routines as strongly disagree, and another 3.75\% as somewhat disagree. This indicates that even within the same population, there is diversity in terms of what users consider to be a reasonable routine.
A reason for this variance is personal preference.
For example, one evaluator said that they needed additional conditions to be stated, without which the routine would be invalid: ``{\em...For the routine if time is Morning, turn the Coffee Maker on, the user should also be at home.}''
In some cases, evaluators did not agree with a particular device being a part of the ``trigger''/``action'', while everything else in the routine seemed reasonable.
These preferences also led to many of the ``somewhat'' ratings.

\subsection{Sequence Generation Study}
\label{sec:sequence_generation}
The sequence generation study measures the ability of \tool to produce valid sequences of lengths larger than 2, based on histories provided by the external evaluators.
We also wanted to test if the validity of the generated sequences improves {\em if we collect home automation sequences from the evaluators and include them in the model}.
That is, we expect that just as NLP-based auto-complete approaches improve in their predictions after learning from the user's data, \tools scenarios for a user will improve once \tool knows  more about the user's home automation context. 
Therefore, prior to the study, we collected user-driven routines and execution indicators from the evaluators and computed home automation sequences for them, using the approach described in Sec.~\ref{sec:data_collection}.
Also, we asked the evaluators to provide us with two histories (\ie sequences of events that would happen in the evaluator's home).
%(\ a set of events that would happen in the morning, after work, or during the day) to use as the histories for generating sequences.

\myparagraph{Methodology} This study was also conducted using an approach similar to the routine comparison study, with a few key differences.
%The sequence generation study was conducted similarly to the routine comparison study with a couple of key differences. 
First, all four model configurations were of the {\em up} flavor, \ie the general {\em up/3-gram} and {\em up/4-gram} models trained on the {\sf HOME} corpus, and specific {\em up/3-gram} and {\em up/4-gram} models for each evaluator trained on the home automation sequences for that evaluator, in addition to the {\sf HOME} corpus.
%Of the four total configurations, two (3- and 4-gram) used a model trained on the {\sf HOME} corpus, and another two (also 3- and 4-gram) used a separate model for each participant, trained on the version of the {\sf HOME} corpus augmented with that participant's home automation sequence.
%Using two separate models that exclude and include the participant's data allowed us to evaluate whether participants found scenarios that were generated by a model that incorporated their own smart home preferences to be more or less reasonable than those simply collected from the crowd.
Second, each of the four models was used to generate 5 successive events, for the two separate histories provided by each evaluator.
The evaluator rated the validity of a generated event based on the history used to generate it, rating 40 events in total.
%Thus, every evaluator rated the validity of 40 individual predictions based on the history, one by one, using the same Likert scale used in the previous study.
We did not use the {\em down} flavor in this study to reduce the cognitive load on the evaluators and prioritize quality responses and rationale by asking no more than 40 questions per evaluator.
%, and because (2) the previous study sufficiently establishes the unnaturalness of DOWN routines, which is likely to extend to longer scenarios as well. 
\input{tables/sequence_validity}

\myparagraph{Results} Table~\ref{tab:sequence-study-results} summarizes the evaluator ratings for the scenarios generated by \tool (\ie 160 ratings per configuration). Evaluators mostly agreed with the routines produced by \tool (\ie over 62\% for the model trained on {\sf HOME} alone (both 3 and 4-gram), and much higher when including the evaluator's data. Further, evaluators justified scenarios as reasonable with additional explanation/situations in most cases and were confident about their choices.
%We now discuss the findings from these results and evaluator comments. 
%, as well as the additional insights from the comments provided by our participants.

\finding{10}{\tool is able to generate reasonable scenarios, {\em despite  previously unseen tokens} in evaluators' histories} Tokens outside the model's vocabulary are known to hurt the predictive capabilities of language models. 
The histories provided by 8 out of 16 evaluators contained tokens that were simply not present in the {\sf HOME} corpus. 
%evaluator's case, the histories provided by the participants were not included in the sequences used to build the {\sf HOME} corpus. 
Despite these disadvantages, \tool delivers a reasonable worst case performance, \ie achieving an overall validity rating of over 62\% in both 3 and 4-gram configurations. 
%This reasonable worst-case performance is a strong indicator of the tremendous potential of this approach. 

% While predicting over unseen tokens is a hard problem, these results indicate the applicability of \tool.

\finding{11}{Including the user's data improves the validity} 
%Providing the model with some sequences from the user significantly boosts the ratings of the generated scenarios 
As shown in Table~\ref{tab:sequence-study-results}, the models including the evaluator's event sequences perform much better, with the overall agreement for the 3-gram and 4-gram models rising to 76.88\% and 67.5\% respectively, from just over 62\% previously.
%This trend is most prominent in the somewhat agree rating, where the 3-gram model with user data outperforms the one without by 12.5\%. 
Note that whether training with or without the evaluator's data, the specific histories provided by the evaluators were mostly outside the scope of the model, \ie only 4 out of the 32 histories provided by our 16 evaluators had subsequences of size 2 or more represented in the training sequences.
This finding indicates that use-cases where end-users may want to use \tool will benefit from this trend, as providing user data will be justified in those cases, and will lead to improved scenarios for end-users.

\finding{12}{Context-shifts may lead to lower ratings} 
%A significant number of the somewhat agree/somewhat disagree ratings occur because of ``context-shifts'', \ie from one smart home scenario to the next} 
\tool can generate event sequences of arbitrary lengths; however, in reality, many scenarios of varying sizes may occur one after the other in the smart home.
We observed that in many cases, users rated the first event of a new contextual scenario very low (\ie somewhat disagree, or somewhat agree), as they could not reason about it in terms of the history.
However, as the new context progressed, the rating improved.
Here is a quote that illustrates this problem (the night-related events happened first, followed by the air quality-related events): ``{\em The first sequence does not make much sense to me because the air quality detection and the security system/night mode are not clearly related, but these events could technically still happen in sequence.  The second sequence makes perfect sense, as you would want the air to be purified if it was detected to be of suboptimal quality.}''.
This finding indicates that \tool generates sequences that may actually span  multiple, discrete contextual scenarios in the home, which may be useful for a longitudinal evaluation of the home.
%, and \tools generates scenarios that may actually encompass multiple scenarios.
%, and for our analysis, we assume the length of the scenario is based on the type of the task (\eg 5 if users are to reason about the events, 10 for policy generation by experts).
%However, from user feedback and trends in the ratings, we noticed that a large number of the ``somewhat'' responses were due to shifts in the context of the smart home scenario, \ie the end of one contextual series of events, and the beginning of another (\eg the end of a morning scenario, followed by the evening scenario). 
%More importantly, we observed that the next event generated after the shift (\ie the supposedly the second event in the new scenario) was generally deemed reasonable by the participant, further bolstering the idea that multiple, discrete contextual scenarios exist in the home, and \tool generates sequences that may actually encompass multiple scenarios.
%For example, consider this quote from a participant describing one such situation (\ie where the night-related events happened first, followed by the air quality-related events): ``{\em The first sequence does not make much sense to me because the air quality detection and the security system/night mode are not clearly related, but these events could technically still happen in sequence.  The second sequence makes perfect sense, as you would want the air to be purified if it was detected to be of suboptimal quality.''}
Moreover, the challenge of capturing these context shifts, and recognizing these scenarios within scenarios, exposes a promising research direction.

\begin{comment}
\begin{figure}[t]
    \centering
    \includegraphics[width=2.5in]{example-image-a} 
    \vspace{-1em}
    \caption{{\small Box-and-whisker results of user ratings for the naturalness of trigger-action routines.\TODO{Add this figure}}}
    \vspace{-1em}
\label{fig:validity-boxplots}
\end{figure}  
\end{comment}

%% file: tables/validity-study.tex
% \usepackage{multirow}
\begin{table}[]
\footnotesize
\caption{{\small A summary of the ratings by evaluators for the routine comparison study. Percentages are out of 160 ratings in total.}}
\label{tab:bigram-study-results}
\vspace{-1em}
\begin{tabular}{c|c|p{0.75cm}|p{1cm}|p{1cm}|p{0.92cm}}
\Xhline{2\arrayrulewidth}
%\rowcolor[HTML]{C0C0C0} 
\multicolumn{6}{c}{\textbf{Routine Study (Bigram Generation)}}                                                                                                                                                                                                            \\ \Xhline{2\arrayrulewidth}
%\rowcolor[HTML]{C0C0C0} 
\multicolumn{1}{c|}{\textbf{Model Config}} & \multicolumn{1}{l|}{\textbf{N-Gram}} & \textbf{Strongly Agree}                 & \textbf{Somewhat Agree}                 & \textbf{Somewhat Disagree} & \textbf{Strongly Disagree}              \\ \hline
                                                                    & 3                                                                  & \textbf{50.63\%} & 19.38\%                                  & 11.87\%                     & 18.13\%                                  \\ \cline{2-6} 
\multirow{-2}{*}{\em up}                                                & 4                                                                  & \textbf{69.38\%} & 13.12\%                                 & 4.38\%                      & 13.12\%                                  \\ \hline
                                                                    & 3                                                                  & 5.0\%                                   & 11.87\%                                 & 15.0\%                     & \textbf{68.13\%} \\ \cline{2-6} 
\multirow{-2}{*}{\em down}                                              & 4                                                                  & 0.63\%                                   & 11.87\%                                  & 19.38\%                     & \textbf{68.13\%} \\ \hline
{\em  none (from survey)}                                                              & \textbf{-}                                                         & \textbf{63.75\%} & 25\%                                    & 3.75\%                      & 7.5\%                                   \\ 
\Xhline{2\arrayrulewidth}
%\rowcolor[HTML]{C0C0C0} 
\end{tabular}
\vspace{-2em}
\end{table}

%% file: tables/sequence_validity.tex
\begin{table}[]
\footnotesize
\caption{{\small A summary of the ratings for the sequence generation study. Percentages are out of 160 ratings in total.}}
\label{tab:sequence-study-results}
\vspace{-1em}
\setlength{\tabcolsep}{3.5pt}
\begin{tabular}{c|c|p{0.75cm}|p{1cm}|p{1cm}|p{0.92cm}}
%\hline
\Xhline{2\arrayrulewidth}
%\rowcolor[HTML]{C0C0C0} 
\multicolumn{6}{c}{%\cellcolor[HTML]{C0C0C0}
\textbf{Event Sequence Generation Study}}                                                                                                                                                                                                              \\ %\hline
\Xhline{2\arrayrulewidth}
%\rowcolor[HTML]{C0C0C0} 
\multicolumn{1}{c|}{\textbf{Training Data}} & \multicolumn{1}{l|}{\cellcolor[HTML]{FFFFFF}\textbf{N-Gram}} & \textbf{Strongly Agree}                 & \textbf{Somewhat Agree}                 & \textbf{Somewhat Disagree} & \textbf{Strongly Disagree}              \\ \hline
                                                                    & 3                                                                  & \textbf{41.25\%} & \textbf{21.25\%} & 15.63\%                     & 21.88\%                                  \\ \cline{2-6} 
\multirow{-2}{*}{{\sf HOME}}                                    & 4                                                                  & \textbf{37.5\%} & \textbf{25.63\%} & 12.5\%                     & 24.38\%                                  \\ \hline
                                                                    & 3                                                                  & \textbf{43.13\%} & \textbf{33.75\%} & 10.63\%                     & 12.5\%                                  \\ \cline{2-6} 
\multirow{-2}{*}{{\sf HOME} + Evaluator's data}                               & 4                                                                  & \textbf{38.75\%} & \textbf{28.75\%} & 16.88\%                     & 15.63\%                                  \\ 
%\hline
\Xhline{2\arrayrulewidth}
\end{tabular}
\vspace{-2em}
\end{table}

%% file: case-studies.tex
\section{Evaluating the Usefulness of \tool (\resq{3})}
\label{sec:case_studies}

We now know that scenarios generated by \tool are drawn from a {\em natural} corpus (\fnumber{6}), and are judged as {\em valid} according to end-users (\fnumber{7}, \fnumber{10}, \fnumber{11}).
This section demonstrates the usefulness of the scenarios generated by \tool. 
We primarily focus on the use of scenarios by security researchers to specify security and safety policies grounded in natural home automation, \ie the motivating example (Sec.~\ref{sec:motivation}).
In addition, we also explore the execution of scenarios by building an execution engine, and provide initial insight into how device manufacturers or platform vendors may benefit from the execution of scenarios.
%Additionally, we also explore the 
%This section validates ``usefulness'' hypothesis through practical use cases.
%, \ie we demonstrate that \tools predictions, derived from natural home automation sequences, are also {\em useful}.
%We examine the use of \tool from the perspective of two key stakeholders: {\sf (1)} {\em security researchers} and {\sf (2)} {\em platform vendors}.
Note that the two use cases we discuss are not the only ways to use \tool; indeed, given \tools generative ability,
%that \tool predicts natural test cases for home automation scenarios, 
the possible use cases for various stakeholders are boundless. 
%Additionally, we develop a few methods/artifacts to enable such evaluations (\eg execution engine, snapshot module, security policies). 
%As we plan to release our system and code upon acceptance, we expect an ecosystem of such modules to develop with community support, for other such use cases.
%\KEVIN{WE seem to lose the notion of security queries and insights that we talk about earlier in the paper. I think it would be good to use these terms to frame our applications/findings.}

%\subsection{Tools To Enable Evaluation}
%For enabling security evaluation using \tools test cases, we develop several artifacts.
%We develop two tools to enable these use cases, and expect an ecosystem of such tools to develop with community support for additional use cases:
We develop two tools to enable these use cases, and expect an ecosystem of additional tools and use cases with community support:

%{\sf (1)} a {\em snapshot module} and {\sf (2)} an {\em execution engine}. 
%We now describe the UP/DOWN test cases predicted by \tools trained model, followed by the {\sf (1)} snapshot module and the {\sf (2)} execution engine.

%\myparagraph{1. UP/DOWN Test Cases} We provide test cases predicted from a trained model, using the UP and DOWN flavors. The UP test cases denote natural event sequences, \ie what would be very likely to happen in a home.
%On the contrary, the test cases created using the DOWN flavor denote highly unnatural event sequences, representing problems that may result from intentional/adversarial misuse.
\myparagraph{1. Snapshot Module} This module tracks the evolution of states of individual devices and the home, as events are executed in the home (\eg the ``locked'' state of the door lock, the home/away mode).
That is, given a scenario, this module provides a snapshot for each event, which shows the holistic state of the home on the event's execution.

\myparagraph{2. Execution Engine} To allow the dynamic execution of the scenarios predicted by \tool, we built an execution engine on top of the SmartThings platform. 
This engine can execute scenarios on real and virtual devices.
% (\ie for simulating situations where a real device may not be available/appropriate).
Our current setup has more than 15 real devices, and can provision an arbitrary number of configurable virtual devices.
%We plan to release these tools, along with \tool, upon publication, and expect an ecosystem of such artifacts to develop with community support, for enabling numerous use cases.

\subsection{Helping Researchers Generate Policies}
\label{sec:security_researchers}
Recall that home automation security policies for prior systems	\cite{cmt18,ctm19,whbg18,nsq+18}  are commonly generated using a use/misuse case security-requirements engineering approach (Sec~\ref{sec:motivation}).
In the example, Alice's key problems were having to imagine all use/misuse cases, which took significant effort, and not knowing whether her policies are indicative of problems that may happen in the wild.
As described in {\em part 2} of the motivating example, using scenarios generated by \tool may address these challenges.
%\tools scenarios may address both these problems, as we described previously in the {\em part 2} of the motivating example (see Section~\ref{sec:motivation}).

\myparagraph{Enacting the motivating example and predicting policies} We had a security researcher (also an author) with prior experience creating smart home policies using the use/misuse case requirements engineering approach, try out \tools approach and scenarios instead.
Our researcher used \tool to generate scenarios in four configurations: the \ {\em up} and {\em down} scenarios described in Sec.~\ref{sec:predicting_test_cases}, as well as two new {\em hybrid} configurations, \ie\ {\em up-down}, which predicts 1-3 {\em down} events for every 10 events in an {\em up} scenario, and {\em down-up}, which does the inverse of {\em up-down}.
We provided histories from 5 additional users (\ie outside the {\sf HOME} corpus) as input for the sequence generation.
%, specifying histories collected from 5 additional users (\ie outside the {\sf HOME} corpus) as input for the sequence generation.
%Additionally, we provided two other {\em hybrid} configurations, \ie {\em up-down}, which predicts 1-3 {\em down} events for every 10 events in an {\em up} scenario, and {\em down-up}, which does the inverse of {\em up-down}. 
% (\ie \xy per user, \xy each using {\em up} and {\em down} flavors).
Each scenario consisted of 13 events, \ie\ 3 events from the history and 10 predicted events.
% and \xy events across \xy histories.

\myparagraph{Results} The researcher analyzed each scenario using the snapshot module, and generated 17 policies from 27 unique violations they detected, spending only a few seconds on scenarios that had no interesting (\ie security/safety sensitive) events, and 1-3 minutes on interesting scenarios, spending about 10 hours. We list all the violations and policies on App.~\ref{app:policies}, and discuss three salient policies: 
%We now discuss three salient policies and the violations that led to them (see Appendix~\ref{app:policies} for the full list).

\begin{itemize}	
\item {\bf Preventing an accidental fire (Pol$_1$):} In one scenario, we discovered that the {\em gas stove} was ON when the user was away. This safety violation could cause a fire, and is addressed with the policy:
%, and is highly probable (\ie as it was predicted using the {\em up} flavor). 
%The following policy addresses this violation: 
{\em the gas stove should be OFF when the user is away.} 
%Note that prior work~\cite{cmt18} also proposes a similar policy, but with other devices (\eg coffee maker). 
\item{\bf Preventing an explosion (Pol$_2$)}
% In a scenario predicted using the {\em down} flavor, 
We discovered that the gas stove is switched ON after the smoke/CO detector has detected smoke, \ie\ {\em when there is already a fire}. A gas leak and fire together could lead to a disastrous explosion, regardless of whether the user is home or away (\ie the violation is unsolvable using Pol$_1$). 
%Note that this is an unlikely scenario (\ie\ {\em down}) (as demonstrated in \fnumber{10}), however, it is possible given a malicious/adversarial threat model (\ie as identified by users \fnumber{11}). We propose the following policy to address this scenario: {\em the gas stove should be OFF when smoke is detected (Pol$_2$).} 
\item{\bf Preventing an accidental privacy violation (Pol$_3$)} We discovered that the indoor security camera would stay ON, even after the user got home, which could lead to a privacy violation. This is a known problem, and vendors (\nest\cite{nestapp}) offer the flexibility of {\em automatically turning the camera OFF when the user is home}
\end{itemize}

We now discuss our core findings from this experiment:

\finding{13}{Scenarios significantly reduce the effort in specifying policies}
The effort involved in developing our policies was far lower than what a fully-manual approach of use/misuse case requirements engineering would have entailed. 
To quote the security researcher from this experiment: ``{\em It's very convenient. The advantage of having the sequences is that they set up a likely or unlikely scenarios without me having to invent it.}''

\finding{14}{All of \tools configurations, \ie\ {\em up, down, up-down and down-up}, are useful} Each of our configurations contributed to the policies generated. In fact, for half of the policies, there is only one configuration that leads to it (as shown in Table~\ref{tbl:policies}). This indicates that the diversity of scenarios created with \tools prediction configurations is valuable.

%Many of the policies resulted from scenarios generated using  Most of our policies were abstracted from multiple unique violations (\ie which could be addressed with the same policy). In some policies, the violations were discovered using multiple configurations (\i

%Some of the policies created were derived from violations discovered usingviolations that were discovered four configuration {\em up-down} and {\em down-up} scenarios resulted in violations that were generally not present in the scenarios generated with solely {\em up}/{\em down}.

%That is, we have already demonstrated that the scenarios generated using the {\em down} flavor are unreasonable/unnatural from the view of the user (\fnumber{10}), and more importantly, that users may consider them unreasonable because they may imply a security/safety hazard (\fnumber{11}).
%Our policy specification revealed additional evidence that the {\em down} flavor generates scenarios that malware may be likely to execute (\eg Pol$_2$), but users or security researchers may find {\em hard to imagine} as reasonable.

\subsection{Detecting flaws in platforms and devices}
\label{sec:platform_vendors}

We explore the benefits of executing scenarios to help platform vendors and device manufacturers evaluate their platforms or partner devices in realistic situations.
On executing a random set of fewer than 10 scenarios in our SmartThings-based execution engine, we discovered three flaws (two platform and one device):
\begin{itemize}
\item {\bf Dropped actions (Flaw$_1$):} On seemingly random occasions, the SmartThings platform was not executing certain events, including security-sensitive events such as locking the door.
\item {\bf Zombie SmartApps (Flaw$_2$):} We observed that routines that were previously ``deleted'' using the \smartthings mobile app~\cite{smartthingsclassic} were persistently executing in the background, and could only be deleted from the Web IDE~\cite{smartthingsIDE}. This can lead to disastrous consequences if the routine were vulnerable, or a malicious SmartApp.
\item {\bf Unsafe Auto-relock default (Flaw$_3$):} 
%Flaw 4 (Device) 
We discovered that the Yale lock~\cite{yale} would stay in the unlocked state after the door closed, unless manually locked from the inside. This {\em auto-relock} feature is standard in regular keypad locks, but implemented as an inconsistent default in most major lock brands, and worse, not presented to the user during device setup with SmartThings.
\end{itemize}

While we independently discovered Flaw$_1$ and Flaw$_2$, they were also previously reported on the \smartthings forum by users~\cite{smartThingsConcurrency, smartThingsZombie}, which further speaks to the ability of our scenarios to uncover problems that naturally occur in end-user homes.
\input{observation_survey}

%% file: related-work.tex
\section{Related Work}
\label{sec:related_work}

%% Home automation security
% 
%This paper is motivated by a gap in the domain of home automation security, and resolves the gap using an approach inspired by prior treatment of software as a natural language.
%This section attempts to address related work in both these domains. 

%\myparagraph{1. Home automation security}
In terms of the analysis of {\em user-driven} routines, the analysis of IFTTT recipes by Surbatovich et al.~\cite{sab+17} is similar to this paper. However, there are key differences.
First, our goal is to generate natural scenarios of home automation, which differs from the objective of prior work (\ie security analysis of individual routines).
Second, due to our focus on {\em home} automation, we primarily study events that take place within the home, unlike prior work, which focuses in IFTTT to study the security/privacy ramifications of connecting external services (\eg email, Twitter) to the smart home.
%  generate scenarios, , and hence, we limit our scope to events that take place within the home, \ie in user-driven routines. 
%In contrast, prior work was intended for exposing security/privacy ramifications of connecting external services (\eg email, Twitter) to the smart home, and hence, focuses on IFTTT recipes.
%Second, the analysis proposed in \tool is {\em predictive}, in contrast to prior work, which relies on static observations.
Third, while prior work examines the safety of individual recipes, \tool holistically explores the event-relationships in home automation. 
%sequences composed from multiple routines.

%Prior work has proposed a diverse set of security systems and analyses for smart homes. 
Further, the natural perspective provided by \tool is complementary to the diverse set of security systems and analyses proposed by prior work.
Consider ContexIoT~\cite{jcw+17}, an access control system that prompts the user for authorization whenever it identifies the use of sensitive operations (\eg unlocking the door) in new contexts.
To measure the frequency of user-prompts (\ie which affect usability), ContexIoT uses random event sequences generated by fuzz testing of IoT apps.
The scenarios predicted by \tool would provide a more realistic input,  
%relative to random event sequences, 
leading to a representative evaluation for such systems.
Similarly, IoTSAN~\cite{nsq+18}, a system that uses model checking to analyze IoT setups for safety, also uses random events for evaluation, and may similarly benefit from \tool. 
Additionally, \tools\ {\em down} flavor can directly contribute to adversarial benchmarks such as IoTBench~\cite{cmt18}, for effective evaluation of future work.
Further, policy-based enforcement for the smart home~(\eg~ProvThings~\cite{whbg18} and IoTGuard~\cite{ctm19}) can significantly benefit from \tools semi-automated approach for creating policies (Sec.~\ref{sec:security_researchers}).

Finally, recent work has also explored the existence and nature of errors in platforms that utilize trigger-action based programming~\cite{Brackenbury:CHI19,Palekar:SafeThings19}, as well as techniques for debugging of these errors~\cite{Corno:CHI19}. 

%Brackenbury \etal~\cite{Brackenbury:CHI19} identify ten classes of TAP bugs and illustrated through a user study that the presence of such bugs made it more difficult for users to interpret the outcome of buggy TAP programs. 
%Palekar \etal~\cite{Palekar:SafeThings19} identify 9 common TAP errors from prior work and develop a framework to classify these errors for the purpose of improving TAP interfaces. 
%Finally, Corno \etal~\cite{Corno:CHI19} present a web interface that enables end users to debug problem that may arise in TAP programs through step-by-step simulation. 
%None of these past studies or approaches attempts to model home automation for the purpose of generating natural event sequences, as \tool does. 
We see our proposed approach as being largely complementary to this body of work, as \tools event sequence generation could lead to the discovery of new bugs/faults not yet considered by researchers, and more advanced or contextualized debugging techniques. 

%lead to %new debugging paradigms, such as suggesting safe routines for ones that has been deemed unsafe or problematic.

%For instance, ProvThings~\cite{whbg18} is a provenance tracker for the SmartThings platform, which can also enforce policies. 
%However, it relies on manually-curated policies that are generated from already-occurred security incidents, or domain expertise. 
%As described in Section~\ref{sec:case_studies}, \tool can help security experts create effective policies arising from natural regularities in user-driven home automation sequences, without significant manual effort.
%Analysis that relies on security or safety policies, such as Soteria~\cite{cmt18} and IoTSan~\cite{nsq+18}, can similarly benefit from \tool. 

%\myparagraph{2. Treating software as a natural language} 

%% file: discussion.tex
\section{Lessons Learned}
\label{sec:discussion}

User-driven home automation is a complex domain, where home automation event sequences {\em created} by users manifest themselves in a highly contextual manner. 
Inspired by prior work in software engineering~\cite{Hindle:ICSE12}, this work is the first to attempt to demystify home automation, through an approach that helps in understanding the non-obvious regularities within such user-created sequences, and generates actionable scenarios.
%The individual aspects of our framework may be iteratively refined in the future. 
%We now discuss the choice of scheduling and the refinements to our execution engine that future work may address.
Next, we discuss key take aways.

 %a \underline{novel research challenge}:%  unique to home automation:
%These arguments motivate a \underline{novel research challenge} that is unique to home automation: 
%\textit{ unlike past research on mobile platforms, characterizing home automation environments solely according to apps published in markets is not sufficient, as users may not necessarily use IoT apps in favor of easily-created user-driven routines}.

\myparagraph{Lesson 1: Home automation is natural, and can be leveraged to generate valid scenarios} Our experiments demonstrate that home automation satisfies the naturalness hypothesis, \ie sequences created by users are implicitly natural, and exhibit logical patterns that make them predictable (\fnumber{6}). 
Moreover, we also demonstrate that statistical language modeling techniques can be used to model home automation and generate reasonably valid sequences (\fnumber{10},\fnumber{11}). 
More importantly, routines generated by \tool are comparable with the routines generated by real users in terms of perceived validity (\fnumber{7}). 
This is an impressive result, given that we used a rather limited {\sf HOME} corpus for generating these routines. 
These results serve as a strong foundation for applying advanced statistical modeling techniques for home automation security, and as we (and other researchers) gather more data from real users, we can expect even better performance from \tool.

%\myparagraph{Lesson 2: Scenarios generated with \tool are valid} Most importantly, the routines generated by \tool are comparable with the routines generated by real users in terms of perceived validity (\fnumber{7}). 
\myparagraph{Lesson 2: \tool can be used to generate useful policies and detect real flaws} \tool generates scenarios that lead to useful policies (\fnumber{13}). 
Moreover, all flavors of \tool prove useful (\fnumber{14}), and our experiments indicate that {\em down} may be inclined towards generating unnatural but unsafe scenarios (\fnumber{8}). 
%like {\em down} are highly useful in generating adversarial scenarios (\fnumber{16}). 
Finally, \tools scenarios have been used to detect two platform and one device flaws, which is another indication of its utility.

\myparagraph{Lesson 3: \tool enables future research opportunities}. \tool may be modified to model temporal information within sequences (\ie the time by which two events should be spaced out when executed), which would allow it to model context-shifts in the home (\fnumber{12}).
Alternately, future work could also extend our execution engine to allow the evaluator to interactively configure the timing between the events within a scenario, at runtime, based on intuition.
Further, future work could modify \tool to enable a more precise perspective, by adjusting the probability distributions of \tools LMs to predict events for a constrained set of devices (\ie those specific to a certain smart home).

\myparagraph{Lesson 4: We need to study more than IoT apps} Our empirical analysis strongly confirms that routines are important for users (\fnumber{1}), and that users express a strong preference for creating their own routines (\fnumber{3}). 
Importantly, we observe a significant mismatch between the available IoT apps and routines created by our users (\fnumber{2}). 
This phenomenon is in complete contrast to app-based platforms such as Android, where users are entirely dependent on apps built by third-party developers.
Thus, user-driven home automation may introduce a set of largely unexplored security implications that may not be apparent from just analyzing IoT apps.
This paper presents an important take away for future research in home automation security: unlike past research in app-based platforms, the methodology of characterizing the environment solely according to apps published in app markets is not viable for characterizing home automation, as users may not use only IoT apps in favor of easily-created routines.
% when trained on sequences that include events from a wide range of devices.

 %\tools scenarios do not contain temporal information (\ie the time by which two events should be spaced out when executed). 
%This 
%Our current engine executes events in the scenarios one at a time, after a gap specified beforehand by the evaluator. 
%Future work may extend our execution engine to allow the tester to interactively configure the timing between test events within a test case, at runtime, based on the testers intuition. 
%Future work may also examine the possibility of adjusting the probability distributions of \tools language models to predict events for a constrained set of devices (\ie those specific to a certain users smart home) when trained on sequences that include events from a wide range of devices.
%Future work may {\sf (1)} extend our execution engine to allow the evaluator to interactively configure the timing between the events within a scenario, at runtime, based on intuition, and {\sf (2)} explore the possibility of adjusting the probability distributions of \tools LMs to predict events for a constrained set of devices (\ie those specific to a certain smart home) when trained on sequences that include events from a wide range of devices.

%% file: survey_methodology.tex
\section{Additional Survey Questions}
\label{app:survey_methodology}

\begin{figure}[t]
    \centering
    \includegraphics[width=2.8in]{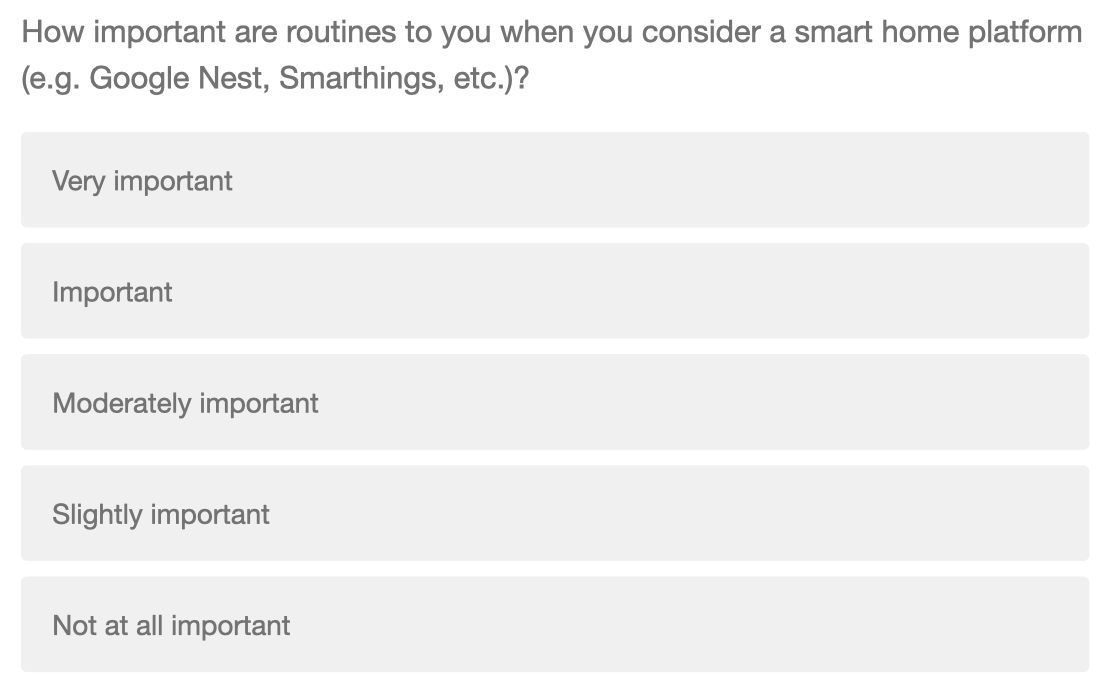} 
	\vspace{-0.2cm}
    \caption{{\small Importance of routines to the users } }
%\vspace{-0.5cm}
\label{fig:how-important}
\end{figure}

\begin{figure}[t]
    \centering
    \includegraphics[width=2.8in]{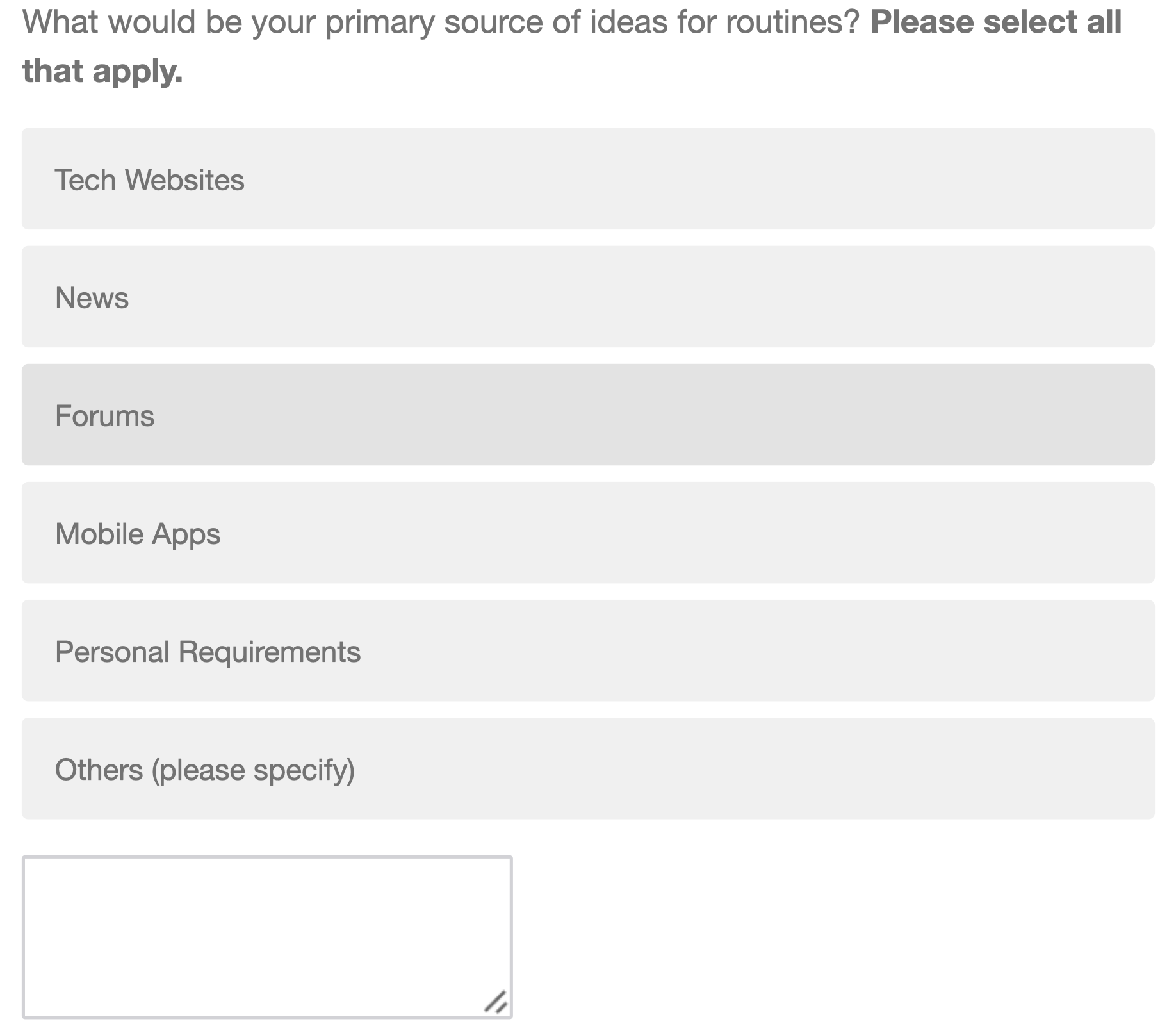} 
	\vspace{-0.2cm}
    \caption{{\small Sources of ideas for routines } }
%\vspace{-0.5cm}
\label{fig:source-ideas}
\end{figure}

\begin{figure}[t]
    \centering
    \includegraphics[width=2.8in]{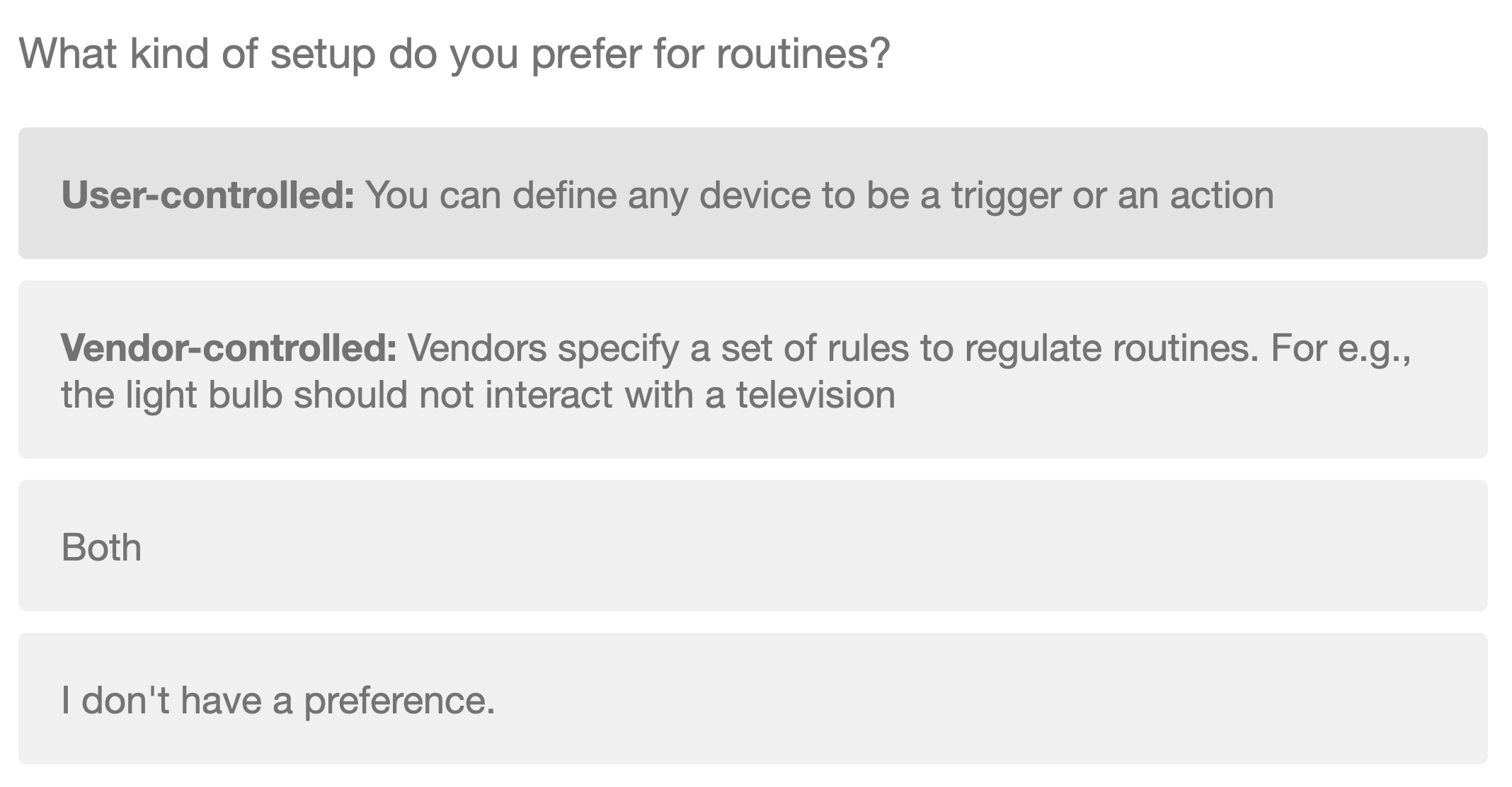} 
	\vspace{-0.2cm}
    \caption{{\small Setup preferred by the users to create routines} }
%\vspace{-0.5cm}
\label{fig:user-controlled}
\end{figure}

\begin{figure}[t]
    \centering
    \includegraphics[width=2.8in]{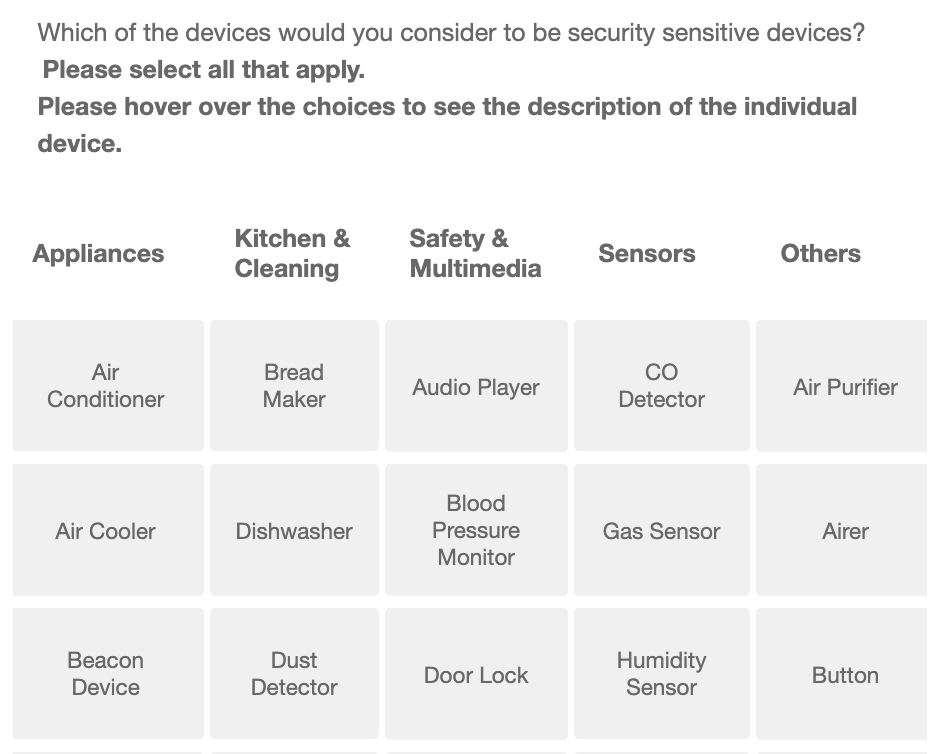} 
	\vspace{-0.2cm}
    \caption{{\small Device-selection screen to select security-sensitive devices} }
%\vspace{-0.5cm}
\label{fig:security-sensitive}
\end{figure}

Aside from collecting routines and execution indicators, we asked users additional questions during the survey, illustrated in Figures~\ref{fig:user-controlled},~\ref{fig:source-ideas},~\ref{fig:how-important}, and~\ref{fig:security-sensitive}.

\section{Survey Instrument for the Routine Comparison and Sequence Generation Studies}
\label{app:user_studies}
\begin{figure}[t]
    \centering
    \includegraphics[width=2.8in]{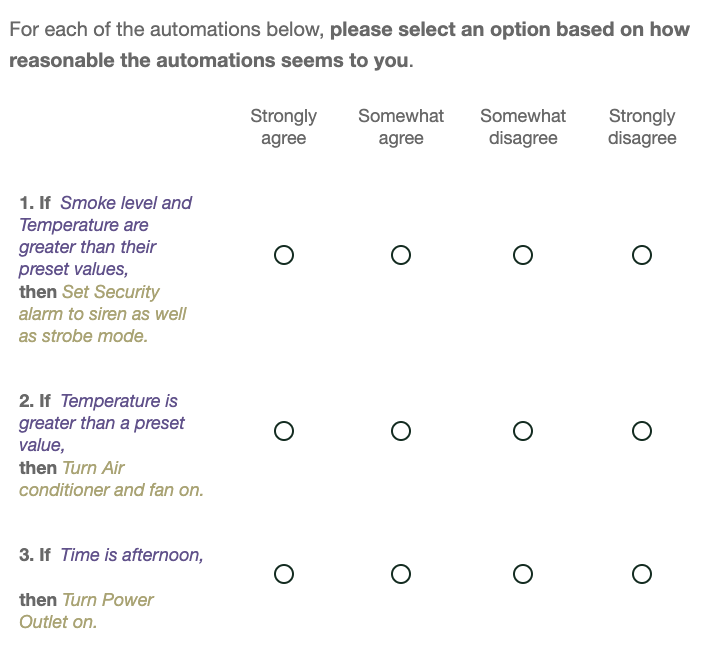} 
	\vspace{-0.2cm}
    \caption{{\small Screen to validate routines} }
%\vspace{-0.5cm}
\label{fig:routine-study}
\end{figure}

\begin{figure}[t]
    \centering
    \includegraphics[width=2.8in]{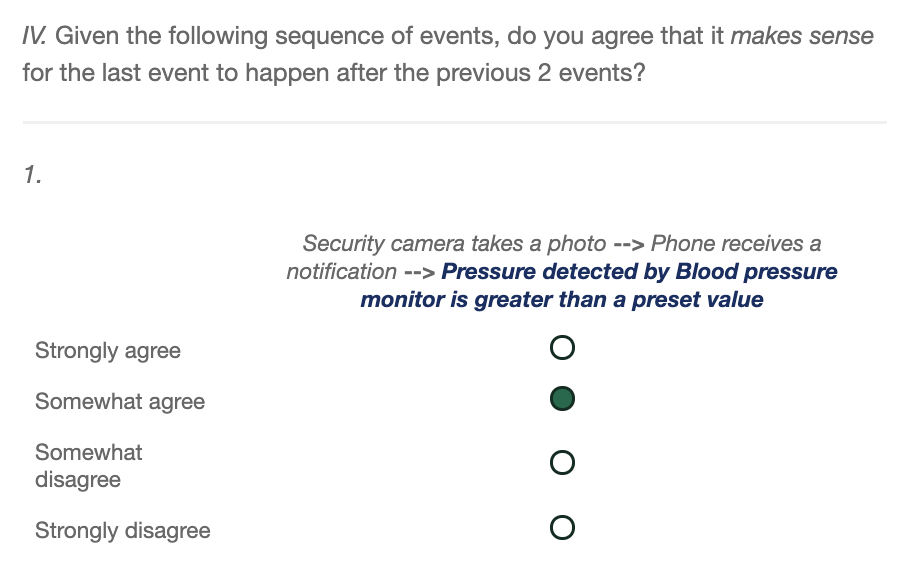} 
	\vspace{-0.2cm}
    \caption{{\small Screen to validate sequences} }
%\vspace{-0.5cm}
\label{fig:sequence-study}
\end{figure}

\begin{figure}[t]
    \centering
    \includegraphics[width=2.8in]{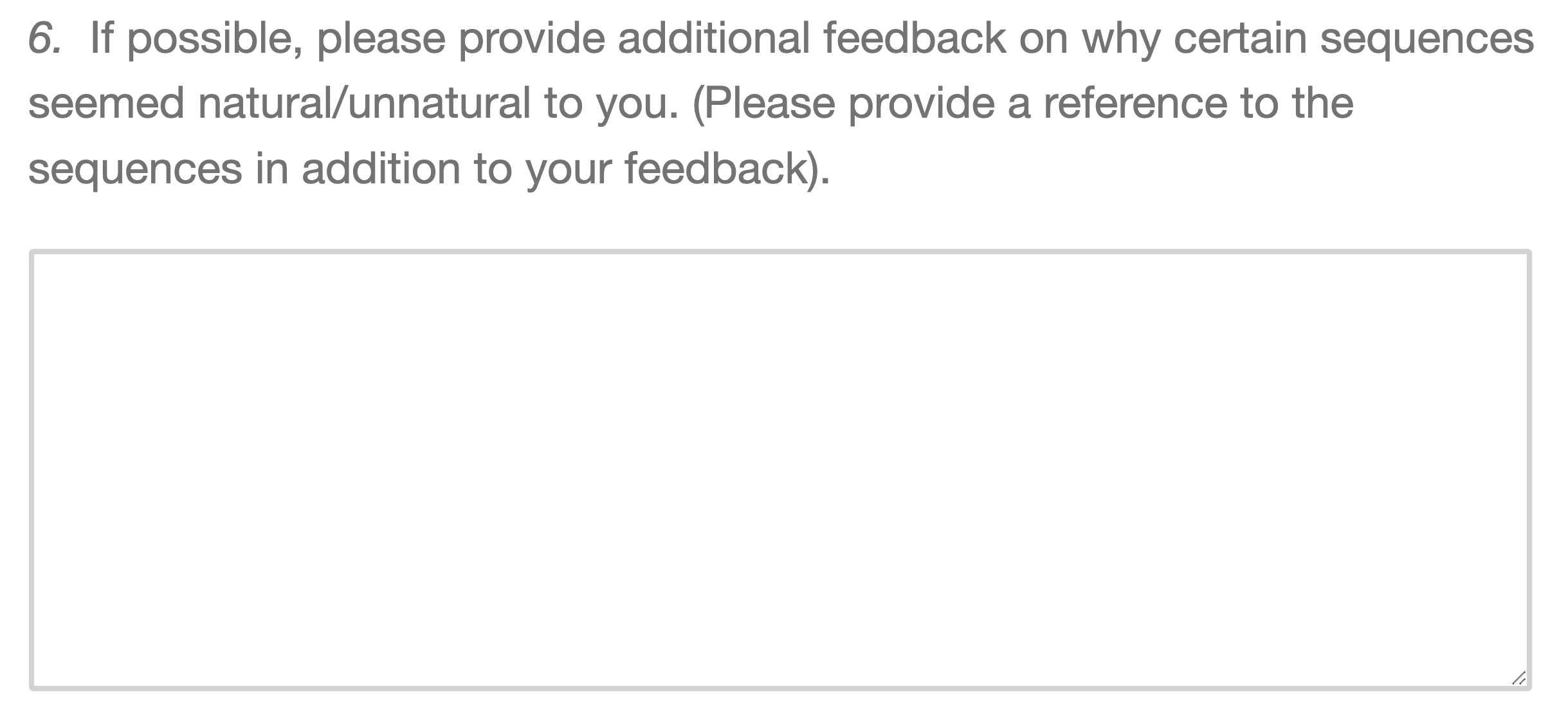} 
	\vspace{-0.2cm}
    \caption{{\small Screen to provide additional feedback } }
%\vspace{-0.5cm}
\label{fig:sequence-study-feedback}
\end{figure}

This section provides the survey instrument for the routine comparison (Sec.~\ref{sec:routine_comparison}) and sequence generation (Sec.~\ref{sec:sequence_generation}) studies. 
Specifically, Figure~\ref{fig:routine-study} illustrates how routines were shown to evaluators in Sec.~\ref{sec:routine_comparison}.
%For each of the 50 routines in the study, evaluators were asked to rate whether they thought it was reasonable to them.
Similarly, Figure~\ref{fig:sequence-study} shows a sequence that one of the evaluators was shown, in Sec.~\ref{sec:sequence_generation}.
%We asked them to rate the sequences based on whether it made sense for the list of events to execute in that order.
Finally, we asked evaluators to provide additional feedback regarding why they rated some routines/sequence as reasonable, as shown in Figure~\ref{fig:sequence-study-feedback}.
%Figure~\ref{fig:sequence-study-feedback} shows the feedback screen for sequence generation study.

% Furthermore, we also associated these devices with its capabilities.
%A capability of a device refers to the function that can be performed using the device.
%Since there are a lot of devices capable of multiple functionalities, we assigned every device to the list possible capabilities (e.g., a Light Bulb can allow users to switch ON and OFF or change its color to WHITE).  
%We collected 178 capabilities from Samsung’s capabilities reference guides~\REF and mapped the devices to these capabilities  based on the use case and the author's intuition.
%Since all these information related to the devices and its functionalities were made available in the survey itself, users were able to spend more time on using these devices to create routines, which is the focus of the survey.

%\TODO{How we performed the survey}
%Finally, instead of asking users to create routines with proper syntax, we asked them to use plain english text.
%This allowed them to be more expressive about their requirements in the smart home.
%We made these design choices to focus on the real end-user requirements and to reduce the cognitive burden on the users where possible.

\begin{comment}
\begin{figure}[t]
    \centering
    \includegraphics[width=2.8in]{figs/sequence_length} 
	\vspace{-0.2cm}
    \caption{{\small 10-fold Cross Validation for Sequence Length} }
%\vspace{-0.5cm}
\label{fig:sequence_length}
\end{figure}
\end{comment}

\input{tables/policies}

\begin{figure}[t]
    \centering
    \includegraphics[width=2.8in]{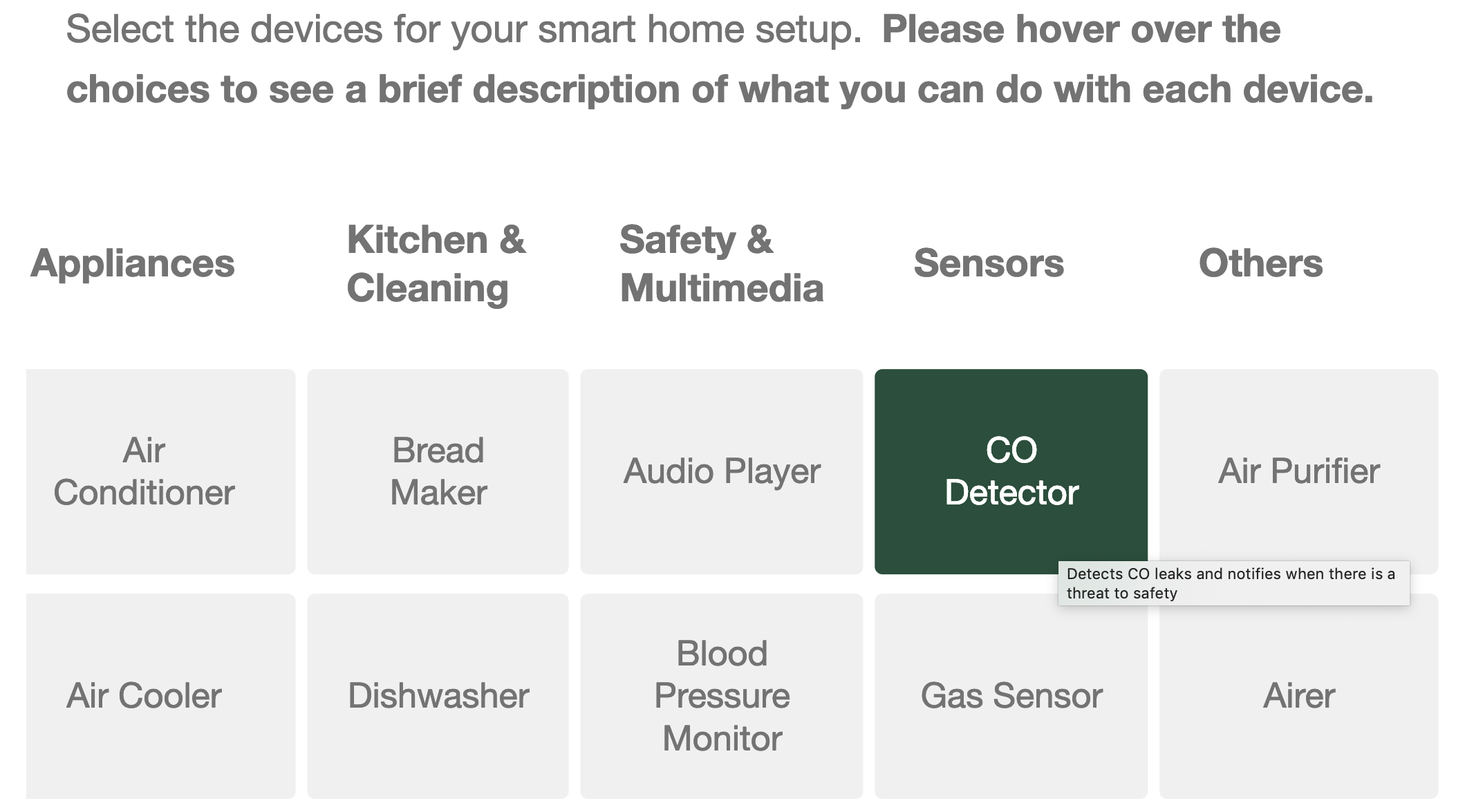} 
	\vspace{-0.2cm}
    \caption{{\small Device Selection Screen} }
%\vspace{-0.3cm}
\label{fig:device-selection}
\end{figure}

\begin{figure}[t]
    \centering
    \includegraphics[width=2.5in]{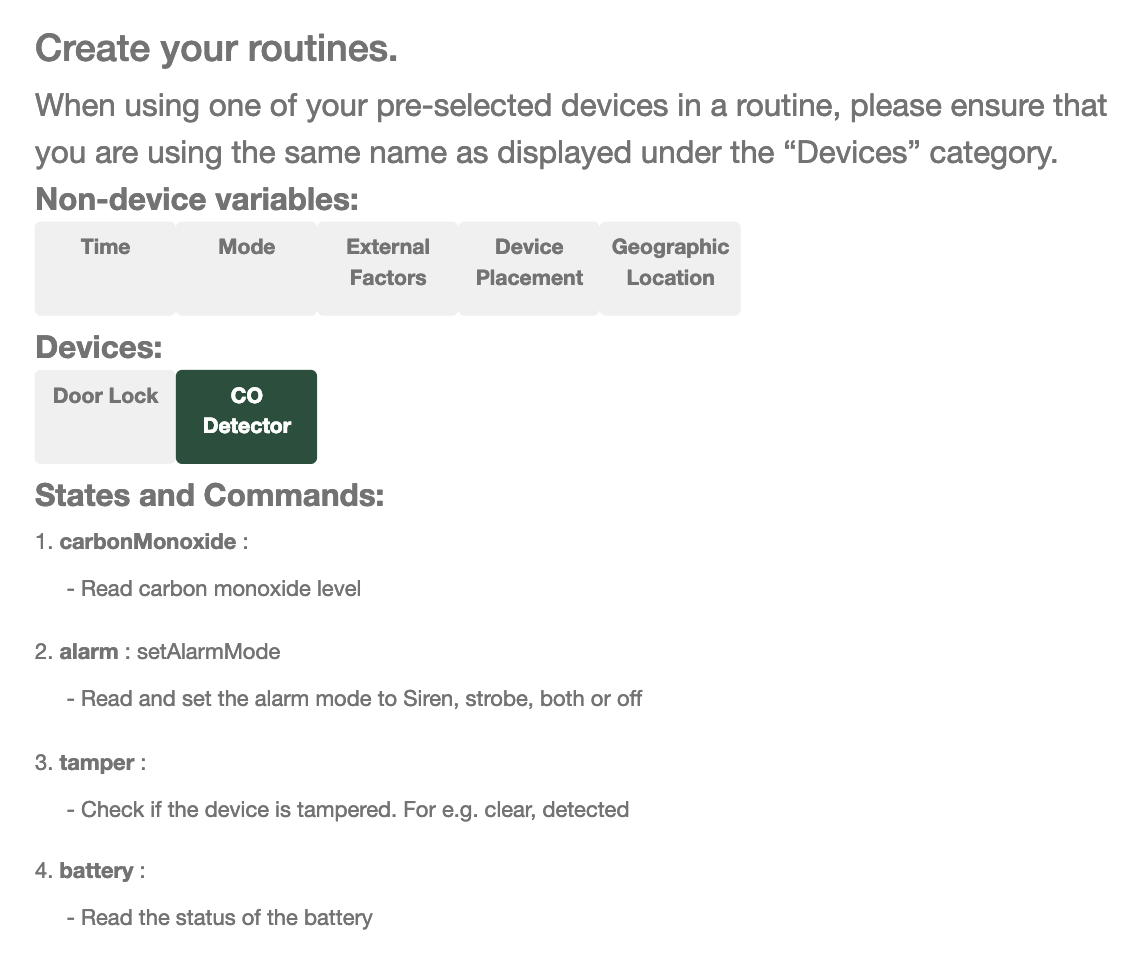} 
	\vspace{-0.2cm}
    \caption{{\small Additional Information about each device}}
\label{fig:attribute}
\end{figure}

\begin{figure}[t]
    \centering
    \includegraphics[width=2.8in]{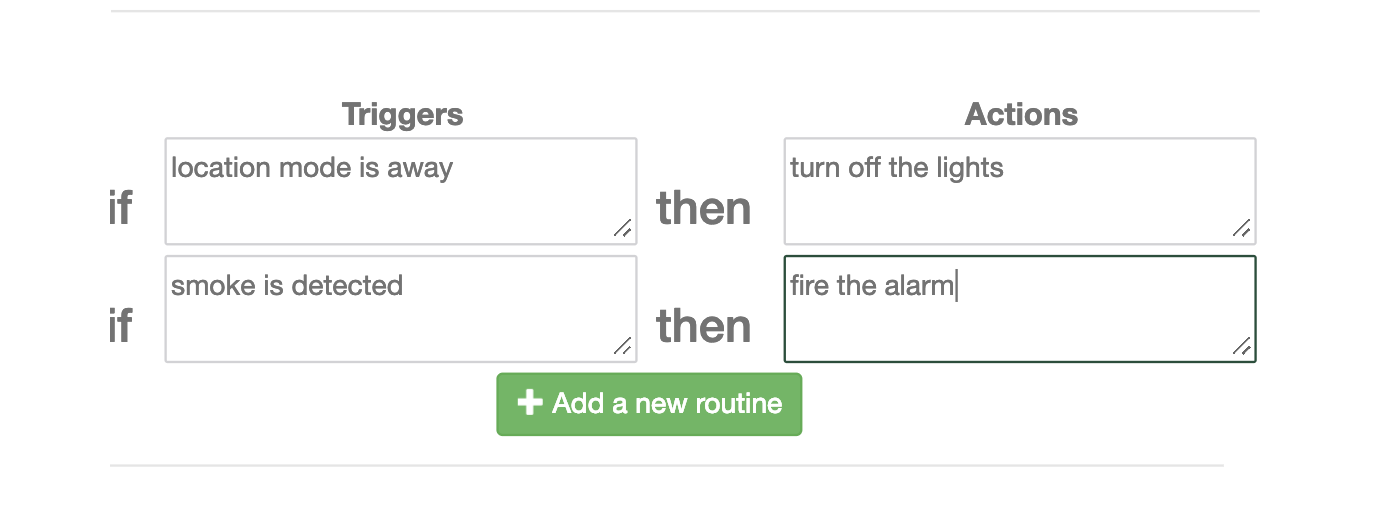} 
	\vspace{-0.2cm}
    \caption{{\small Routine Creation} }
%\vspace{-0.5cm}
\label{fig:routines_creation}
\end{figure}

%%%%%

\begin{figure}[t]
    \centering
    \includegraphics[width=2.8in]{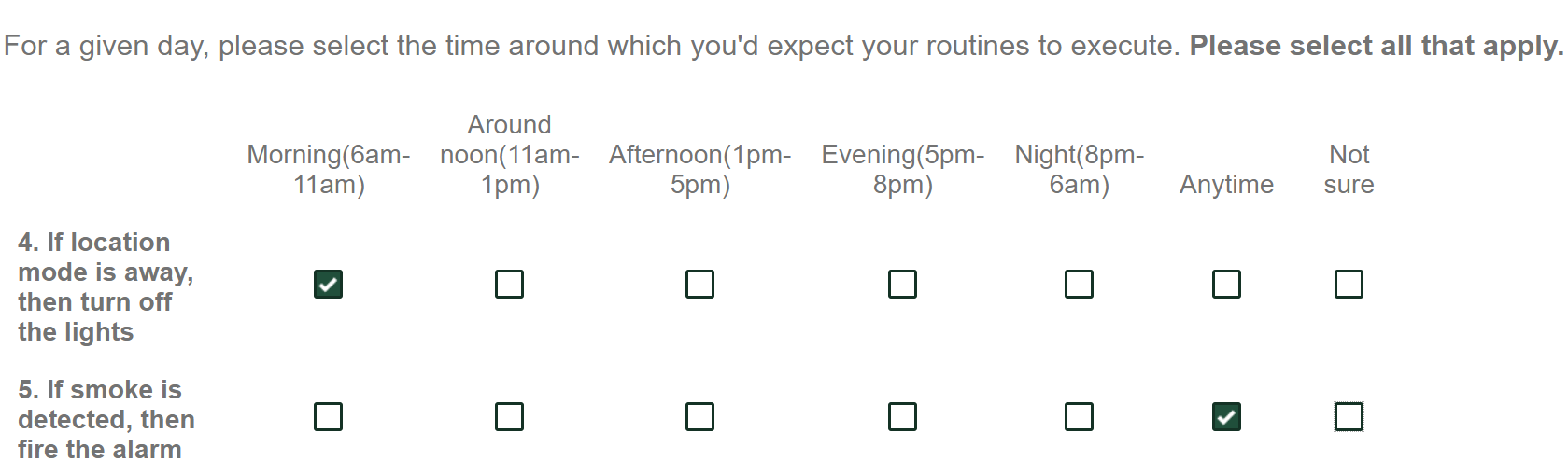} 
	\vspace{-0.2cm}
    \caption{{\small Execution Indicator: Time of the Day} }
%\vspace{-0.5cm}
\label{fig:day}
\end{figure}

\begin{figure}[t]
    \centering
    \includegraphics[width=2.8in]{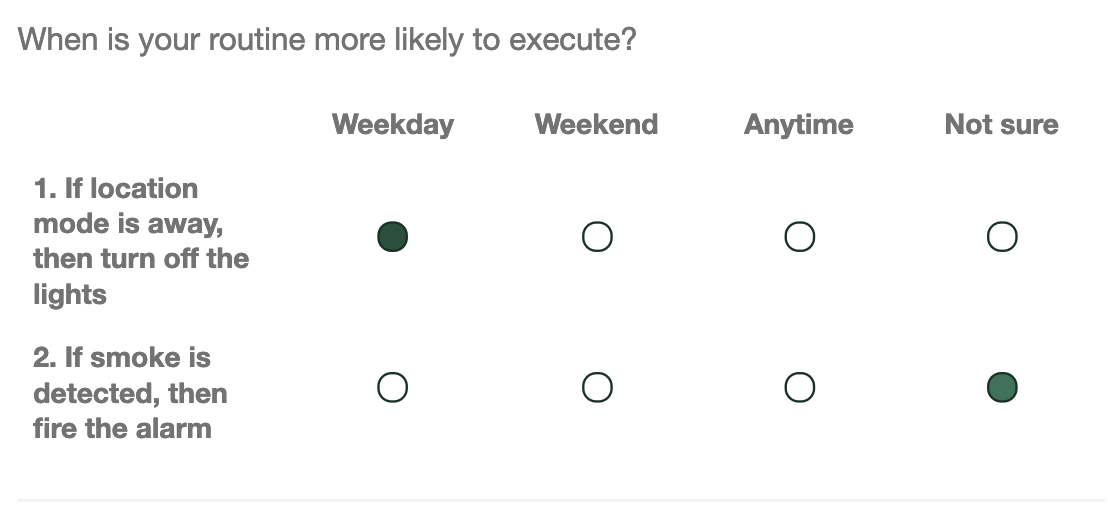} 
	\vspace{-0.2cm}
    \caption{{\small Execution Indicator: Time of the week} }
\label{fig:week}
\end{figure}

\begin{figure}[t]
    \centering
    \includegraphics[width=2.8in]{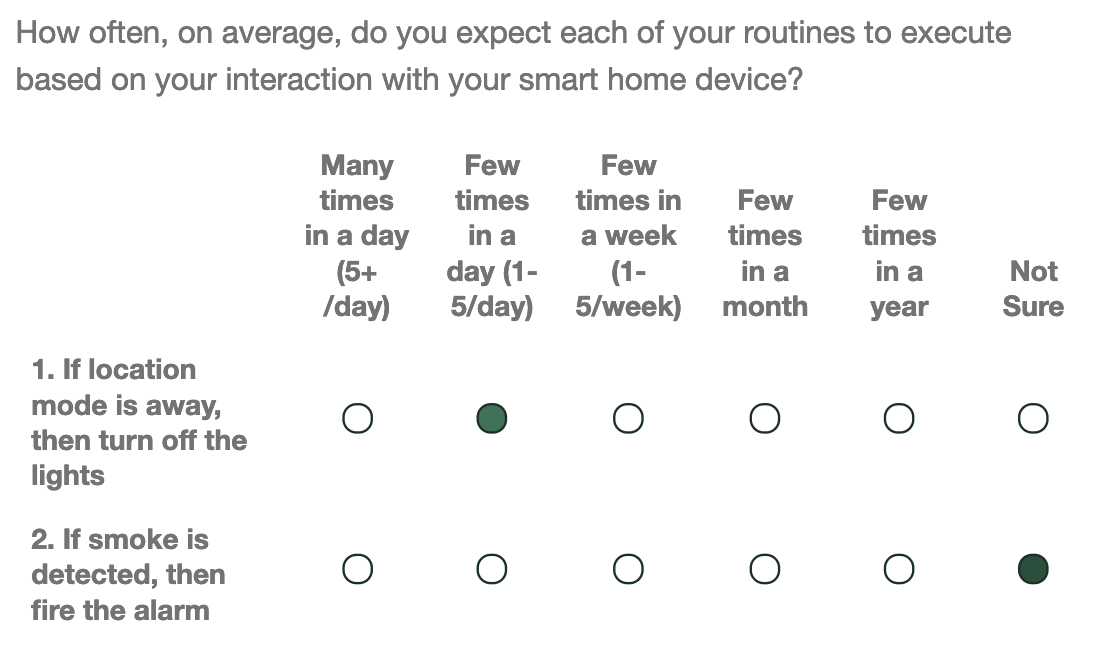} 
	\vspace{-0.2cm}
    \caption{{\small Execution Indicator: Frequency} }
%\vspace{-0.5cm}
\label{fig:frequency}
\end{figure}

\begin{comment}
%%ADWAIT: We never talk about this in the design. This is not an execution indicator, per se.
\begin{figure}[t]
    \centering
    \includegraphics[width=2.8in]{figs/appendix/manual-scheduled} 
	\vspace{-0.2cm}
    \caption{{\small Execution Indicator: Routine Category} }
%\vspace{-0.5cm}
\label{fig:category}
\end{figure}
\end{comment}

%\begin{itemize}
%\item {\em How important are routines to you?(Figure~\ref{fig:})}
%\item {\em What would be your primary source of ideas for routines?}
%\item {\em What kind of setup do you prefer for routines?}
%\end{itemize}

%This allowed  to be more expressive about their requirements in the smart home.
 
%\subsubsection{Additional Questions:}
%Finally, we asked additional questions related to: 
%{\sf (1)} the primary sources for the ideas to create routines Figure~\ref{fig:}, {\sf (2)} how important they think the routines are when they consider the smart home platform, {\sf (3)} Do they consider security while creating these routines Figure~\ref{fig:}, and {\sf (4)} if they trust the routines created by third party developers Figure~\ref{fig:}. 

%% file: tables/policies.tex
\begin{table*}[t]
\centering
\scriptsize
\caption{{\small List of smart home violations, the core problem found by the violations, and the corresponding policies developed from them.}}
\label{tbl:policies}
\setlength{\tabcolsep}{2pt}
\begin{tabular}{p{0.3\textwidth}|p{0.28\textwidth}|p{0.3\textwidth}|p{0.12\textwidth}}
  \Xhline{2\arrayrulewidth}
 {\bf Violation}  & {\bf Problem} & {\bf Policy} & {\bf Flavor}\\ 
 
  \Xhline{2\arrayrulewidth}  
   \Xhline{0.25\arrayrulewidth}
   - Gas stove on when user away. \newline - Gas stove is on when user is on vacation. & Gas stove is on when user is not at home. & {\bf (Pol$\bf _1$)} Gas stove should not be on when the user isn't home. & {\em down, up-down, down-up} \\
   
     \Xhline{0.25\arrayrulewidth}
    - Gas stove is on when smoke has been detected previously. & Gas stove is on after smoke detector detects smoke. &{\bf (Pol$\bf _2$)} Gas stove should not be on when smoke is detected. & {\em up} \\
    
  \Xhline{0.25\arrayrulewidth}
  - Security camera is taking images when motion is detected. However, the user is home, which justifies motion as well as potentially violates user's privacy. \newline - Security camera on when user is home. \newline - Security camera is on and taking images when presence sensor detects that user is present at home & Security camera is on when the user is at home. & {\bf (Pol$\bf _3$)} Security camera should be off when the user is home to preserve privacy. & {\em up, down, up-down, down-up} \\
      
  \Xhline{0.25\arrayrulewidth}
  - Door is opened when user is away, but the user doesn't receive a notification. \newline - Door is opened when user is in vacation, but user isn't notified. \newline - Door is opened but the presence sensor detects that user isn't present at home and user is not notified. & Door sensor senses that the door is open when user is not at home. & {\bf (Pol$\bf _4$)} User should be notified if the door sensor senses the door is open when the user is away. & {\em up, up-down} \\
  
  \Xhline{0.25\arrayrulewidth}
  - Window opened when user away, but the user isn't notified. & Contact sensor senses that the window is open when user is not at home. & {\bf (Pol$\bf _5$)} User should be notified if the contact sensor senses the window is open when the user is away. & {\em up, down} \\
  
  \Xhline{0.25\arrayrulewidth}
  - Water valve closed when fire sprinkler on. & Water valve is closed when fire sprinkler turns on. & {\bf (Pol$\bf _6$)} Water valve should not be closed if the fire sprinkler is on as a reaction to fire/smoke. & {\em down, down-up} \\
  
  \Xhline{0.25\arrayrulewidth}  
  - Air purifier turns off when CO is detected & Air purifier turns off when CO detector detects CO. & {\bf (Pol$\bf _7$)} Air purifier should not turn off automatically if CO is detected. & {\em down} \\
    
    \Xhline{0.25\arrayrulewidth}
	 - Shades/Blinds are opened in the morning, but the user is away. \newline - Shades/Blinds opened in vacation mode. & Shades/Blinds open when user is away. & {\bf (Pol$\bf _8$)} Window shades should not open when user is not home. & {\em up, down, down-up} \\
    
    \Xhline{0.25\arrayrulewidth}
    - Door is unlocked after gas level is detected to be high or alarm going off, supposedly as a safety measure, but user is away. \newline - Door unlocked in vacation mode. \newline - Door is unlocked state but presence sensor is detecting that user isn't present at home. & Door is unlocked when user is not at home. & {\bf (Pol$\bf _9$)} User should be prompted before the door is unlocked automatically for any reason, when user is away. & {\em down, up-down, down-up} \\
    
    \Xhline{0.25\arrayrulewidth}
        - Door stays unlocked even after user leaves home. & Door lock stays unlocked when user leaves home. & {\bf (Pol$\bf _{10}$)} Door should lock automatically when mode changes from home to away. & {\em down} \\
        
    \Xhline{0.25\arrayrulewidth}
  - Door remains unlocked when the sleep monitor detects that the user is sleeping. & Door is unlocked when user is sleeping. & {\bf (Pol$\bf _{11}$)} Door should be locked when the bedroom sleep monitor detects that user is sleeping to ensure safety. & {\em up-down} \\
  
   \Xhline{0.25\arrayrulewidth}
        - The sleeping monitor detects the user as sleeping, and the garage door is open. & Garage door is open when user is sleeping. & {\bf (Pol$\bf _{12}$)} Garage door should be closed when bedroom's sleep monitor detects that user is sleeping. & {\em up-down} \\ 
        
    \Xhline{0.25\arrayrulewidth}
        - The induction cooktop is on after the sleep monitor detects that the user is sleeping. & Electric appliance which is a potential fire hazard is on when user is sleeping. &  {\bf (Pol$\bf _{13}$)} Induction cooktop should not be on when user is sleeping. & {\em up-down} \\
     
      \Xhline{0.25\arrayrulewidth}
        - Garage door opened when user is away. \newline - Garage door opened when user in vacation mode. & Garage door is open when user is not at home. & {\bf (Pol$\bf _{14}$)} Garage door should be closed when user is not home. & {\em down, up-down, down-up} \\
        
     \Xhline{0.25\arrayrulewidth}
        - Glass break is detected when user is away but user is not notified. \newline - Glass break is detected in vacation mode but user isn't notified. & Glass break is detected but the user is not notified. & {\bf (Pol$\bf _{15}$)} User should be notified when glass break is detected. & {\em down-up} \\
        
     \Xhline{0.25\arrayrulewidth}
        - Security alarm turned off when smoke is detected and user is away. & Security system turned off when smoke detector detects smoke. & {\bf (Pol$\bf _{16}$)} Security alarm should not be off when user isn't home. & {\em down-up} \\ 
        
     \Xhline{0.25\arrayrulewidth}
        - Fire sprinkler on when there is no fire. & Fire sprinkler is on for no reason. & {\bf (Pol$\bf _{17}$)} Fire sprinkler should only be on when there's fire detected in the home. & {\em up-down, down-up} \\          
  \Xhline{2\arrayrulewidth}
  \end{tabular} 
%\vspace{-0.2cm}
\end{table*}

%% file: policies.tex
\newpage
\section{Policies generated using \tool}
\label{app:policies}

Table~\ref{tbl:policies} provides the security/safety policies generated using \tool.